# Diamond-to-graphite transformation under hypersonic impact

Abhijit Biswas,[a,*] Aniket Mote,[b,#] Rajib Sahu,[c,#] Marcelo Lopes Pereira Junior,[a,d,#] Shuo Yang,[e] Sudaice Kazibwe,[f] Jishnu Murukeshan,[a] Raphael Benjamim de Oliveira,[a] Guilherme da Silva Lopes Fabris,[g] Shreyasi Chattopadhyay,[a] Gelu Costin,[h] Jianhua Li,[i] Robert Vajtai,[a] Ching-Wu Chu,[f] Lizhong Lang,[j] Yu Zou,[j] Liangzi Deng,[f] Tobin Filleter,[e] Douglas Soares Galvão,[g] Christian Kübel,[c,k] Thomas E Lacy Jr,[b] and Pulickel M. Ajayan,[a,*]

**AFFILIATIONS**

[a]Department of Materials Science and Nanoengineering, Rice University, Houston, TX, 77005, USA

[b]J. Mike Walker '66 Department of Mechanical Engineering, Texas A&M University, College Station, TX 77843, USA

[c]Karlsruhe Nano Micro Facility (KNMFi) & Institute of Nanotechnology (INT), Karlsruhe Institute of Technology (KIT), Kaiserstr. 12, 76131 Karlsruhe, Germany

[d]College of Technology, University of Brasília, Brasília, DF, 70910-900, Brazil

[e]Department of Mechanical & Industrial Engineering, University of Toronto, 5 King's College Road, Toronto, M5S 3G8, Canada

[f]Department of Physics and Texas Center for Superconductivity (TcSUH), University of Houston, Houston, TX, 77004, USA

[g]Applied Physics Department and Center for Computational Engineering & Sciences, State University of Campinas, Campinas, SP, 13083-970, Brazil

[h]Department of Earth, Environmental and Planetary Sciences, Rice University, Houston, TX, 77005, USA

[i]Shared Equipment Authority, Rice University, Houston, TX, 77005, USA

[j]Department of Materials Science and Engineering, University of Toronto, 5 King's College Road, Toronto, M5S 3G8, Canada

[k]Institute of Materials Science, Technical University Darmstadt (TUDa), Peter-Grünberg-Str. 2, 64287 Darmstadt, Germany

[*]Corresponding authors: **abhijit.biswas@rice.edu**; **ajayan@rice.edu**

[#]A. Mote, R. Sahu, and M. L. Pereira Junior contributed equally to this work.

## Abstract

Consolidation of diamond particles into high-strength bulk materials and composites is extremely challenging as it requires high sintering temperatures, at which diamond transforms into graphite, making it impossible to retain the sintered diamond phase. Here we report a synthetic approach based on spark plasma sintering that allows the stabilization of micron-scale diamond grains in composites, using cubic boron nitride (cBN) as the matrix and cobalt (Co) particles as a stabilizer, resulting in a hard-to-machine, tough material. We performed hypersonic speed impact tests on the composites using metal projectiles of different sizes. The composite withstands impact from ∅1 mm metal projectiles (speed Mach 7.5) but breaks apart during the impact from a larger ∅4 mm metal projectile (speed Mach 8.45). Interestingly, during this microsecond-scale impact and fracture event, embedded diamond particles in the composite undergo near-complete phase transformation to graphite. In-depth micro-structural characterizations of the fractured composite, supported by the molecular dynamics simulations reveal details on the impact-induced phase transformation and the creation of diamond-graphite interfaces, suggesting that the energy absorption is primarily enabled via phase change of diamond. Our findings provide a pathway to stabilize the diamond phase and a new understanding of phase transformation of diamond under extreme conditions.

## Introduction

Diamond, the hardest known material, possesses exceptional mechanical [1,2], and thermal properties [3], making it indispensable for extreme condition applications [4,5]. For advanced industrial use, scalable production of diamond and diamond-based composites is necessary [6–8] and large-scale synthetic polycrystalline diamond is mostly achieved through vapor phase deposition of diamond coatings or by high pressure, high temperature methods [6,9–16]. Starting with the early seminal works by Bundy, Hall and co-workers [9,13,17] on successful synthesis of diamond in the laboratory, enormous interest has led to the synthesis of diamond [10,16], given its huge application potential [5,12]. Although diamond has been produced from the sintering of micron size powders [13,17], as well as using metal binders [18,19], production of graphite-free large-scale sintered diamond at lower pressures still remains challenging [12,16], and a topic of strong interest. In general, below several giga-pascal of pressure, high temperature sintering creates graphitization of the diamond and the product normally is a mixture of the two phases (diamond and graphite) [16,20,21]. If indeed low pressure (lower than typically used in diamond synthesis) sintering of diamond without graphitization can be achieved, this will be an enormous achievement.

One approach, as yet unexplored, for the synthesis of diamond or diamond-based composites at lower pressure would be to use structurally and functionally compatible materials, such as cubic boron nitride (cBN) as the matrix and well-known stabilizers for diamond synthesis such as transition-metals [18,19,22–25] to create multi-phase composites. cBN also has similar sintering characteristics as diamond, with strong affinity to transform to the hexagonal phase during low pressure high temperature sintering [26]. However, during sintering, cBN could act as a heat-sink [27], lowering the effective local temperature by dissipating heat, as well as reduce the surface dangling bonds of diamond by creating robust C-N and C-B bonds and hence the graphitization process. Simultaneously, the transition-metal cobalt (Co) acting as a catalyst [28] could prevent the formation of graphitic $sp^2$-phase, thus stabilizing the inherent micro-structure of diamond. In our work reported here we present the hypothesis that the three-phase composition could be used as a recipe to retain the cubic phase during sintering and experimentally demonstrate this via spark plasma sintering (SPS). This result is a unique demonstration of how a diamond-cBN composite can be created via sintering of their micropowder precursor without the transformation into their respective hexagonal phases.

Concurrently, the diamond-to-graphite [29] (or vice-versa [30]) transformation embodies one of the most intriguing phase transitions; thus, understanding and controlling this phase transformation is fundamentally and technologically important [31]. Traditionally, the diamond-to-graphite phase transformation [31–40] has been observed through various temperature-pressure driven pathways and exploring new conditions, where diamond-to-graphite phase transformation can be fundamentally different [41–43], and is exciting. Diamond-to-graphite (or vice-versa) phase transformation typically proceeds via extensive bond breaking and rehybridization [31] rather than diffusion-less shear-dominated pathways [42]. The crystallographic mismatch between the cubic lattice of $sp^3$ diamond and the hexagonal lattice of $sp^2$ graphite makes the diffusion-less, shear-dominated pathway quite improbable, unless the system is subjected to extreme conditions [44], such as mechanical impact.

Here, we report the stabilization of large-scale hard-to-machine, tough diamond-cBN-Co composites from their respective micron-size grain powders using SPS. We show that under impact using projectiles at hypersonic speeds (>Mach 8) (closer to hypervelocity where shock-wave effects become prominent), the embedded diamond particles in the composites can transform to graphite in microseconds. Comprehensive experimental analyses and molecular dynamics simulations of the impact events provide insights into diamond's phase stability and atomic-scale transformations that happen under extreme conditions. Our findings could provide a fundamental new understanding of diamond-to-graphite transformation under hypersonic impact, a topic that is becoming increasingly important. Furthermore, it could enable microstructural engineering of diamond-based multi-phase composites [45–48], with potential applications for extreme condition technologies [44,49,50].

## Results and Discussion

### *Synthesis of diamond-cBN-Co composites*

We synthesized diamond-cBN-Co composites (equal wt%) using spark plasma sintering (SPS) (see **Methods** and Supplementary material **Fig. S1**) at 1400 °C and 90 MPa pressure. The starting materials are diamond powders (particle size ~50-100 μm), cBN powders (particle size ~0.5-1 μm), and Co powders (particle size ~1-2 μm). The cBN and Co particles were added to prevent the $sp^2$ bond formation and graphitization of diamond [10]. The judicious choice of sintering temperature depends on the fact that Co has a melting point of 1495 °C (1768 K). Other SPS synthesis efforts to obtain diamond-based composites either without using Co or cBN or using smaller (larger) grain sizes of diamond (cBN) show less success (producing either more graphitic phases or non-sintered soft disks (**Figs. S2, S3** and **S4**) [51]. After the optimized sintering attempt, we obtained a dark colored disk (inset of **Fig. 1a**) with a measured density of ~3.32 g/cm$^3$ (by using the solid cylindrical method). Structurally, X-ray diffraction (XRD) shows diamond D (111) and cBN (111) Bragg peaks dominating (**Figs. 1a** and **S1**), their respective most energetically stable facets. We could also observe very small peaks corresponding to graphite (G), suggesting that small amounts of graphite could also form during sintering. Interestingly, although the sintering temperature exceeds the onset temperature of diamond graphitization, yet we only observed a small fraction of graphite G-peak. Also, without using cBN, composite made of diamond and Co produces more graphitization (**Fig. S4**). This indicates that possibly B and N from cBN chemically bind to the exposed surface C atoms of diamond, creating robust C-N and C-B bonds and reducing the surface dangling bonds and hence the graphitization process. We also investigated the orientation of the crystal planes of diamond. The pole figure map shows predominantly (111) faceted diamond grains in random orientation throughout the bulk (**Figs. 1b** and **S5**). Non-destructive bulk X-ray microscopy (XRM) imaging further exhibits randomly oriented diamond grains (red colors) throughout the bulk of the composite (**Fig. 1c** and Supporting movie **S1**). As shown through XRM, the diamond particles are found to be less dense and much smaller in size when a diamond-Co disk was attempted to sinter without using cBN (**Fig. S6** and **movie S2**), showing the importance of cBN as a matrix [52]. The three-phase composite shows $sp^3$-hybridized bonding (e.g. C-N and C-B) as evident through X-ray photoelectron spectroscopy (XPS) (**Fig. S7**), that provides an improved interfacial thermal stability between cBN and diamond phase [52,53].

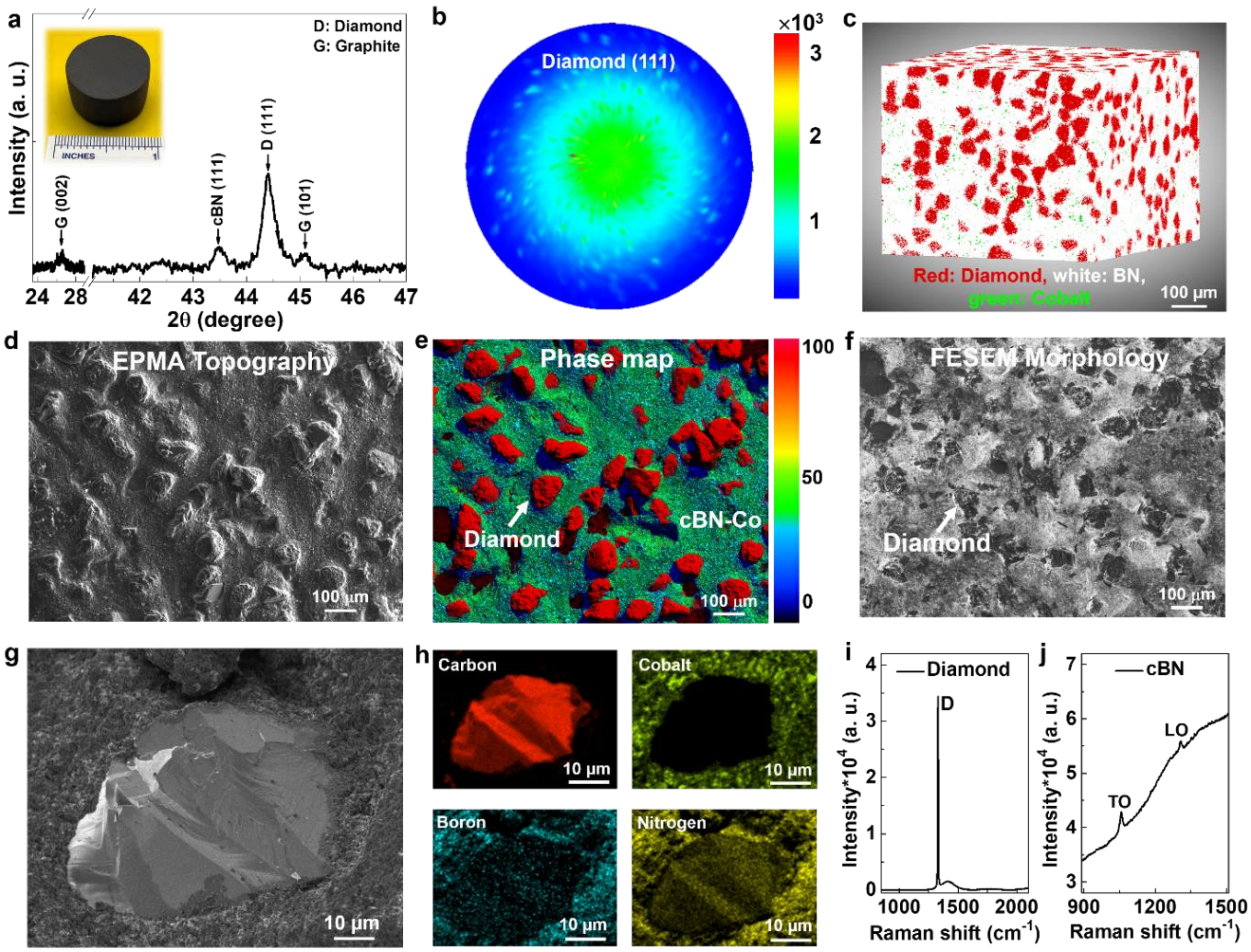

**Fig. 1. Characterizations of the diamond-cBN-Co composite. (a)** XRD of the composite shows Bragg peaks of diamond (D), cBN, and graphite (G). Inset shows a spark-plasma sintered black colored disk. **(b)** XRD pole figure map of (111) diamond peak shows the randomly oriented intensity profiles from the entire sample. **(c)** Three-dimensional X-ray microscopy imaging shows the diamond particles (red) throughout the composite. **(d)** EPMA secondary electron image showing the topography, where diamond crystals stand out as high relief grains in the cBN-Co matrix. **(e)** EMPA WDS phase map shows diamond (red), cBN (green), and Co (light blue). **(f)** FESEM shows the presence of diamond particles with sizes ~50-100 μm (black). **(g), (h)** A diamond particle is embedded in the cBN-Co matrix. Along with diamond, cBN and Co-rich regions with much smaller particle sizes (~1-2 μm) also exist. **(i)** Raman spectroscopy shows D-band (~1330.9 $cm^{-1}$) phonon mode from diamond particle. **(j)** Raman spectroscopy in the non-diamond regions shows transverse optical (TO ~1057.08 $cm^{-1}$) and longitudinal optical (LO ~1307.64 $cm^{-1}$) phonon modes, characteristics of cBN.

The electron probe micro-analysis (EPMA) secondary electron image (SEI) and its wavelength dispersive spectrometry (WDS) elemental maps combining all the elements (**Figs. 1d** and **1e**) also shows the surface topography and presence of particles of diamond (red), while the rest of the regions are from uniformly distributed cBN and Co particles, with an obtained wt% of 34.58% (Carbon), 34.54% (BN), and 30.88% (Co). The field emission scanning electron microscopy (FESEM) (**Fig. 1f**) shows particle morphology, further confirming the presence of diamond grains (sizes ~50-100 μm). We also show the FESEM image of a single diamond particle (**Fig. 1g**). Its energy dispersive X-ray spectrometry (EDS) elemental map shows that it's a diamond particle embedded with cBN and Co particles (**Fig. 1h**). Raman spectroscopy and map from these grains shows the characteristic diamond D-band peak at ~1330.9 $cm^{-1}$ (**Fig. 1i** and **Fig. S8**) [12]. Raman spectroscopy on non-diamond regions show the transverse optical (TO ~1057.08 $cm^{-1}$) and longitudinal optical (LO ~1307.64 $cm^{-1}$) phonon modes (**Fig. 1j**), confirming the presence of cBN particles [27]. The composite was found to be extremely hard-to-machine (**Fig. S9**) and electrically conducting (resistivity ~217 mΩ-cm) at room temperature with a linear V-shape magneto-resistance (due to metallic Co particles embedded in the insulating matrix) (**Fig. S10a**) [54]. In addition, the composite shows a ferromagnetic behavior at room temperature (**Fig. S10b**).

### *Hypersonic (near-hypervelocity) speed impact testing of the composite*

Two diamond-cBN-Co composite samples were subjected to hypersonic speed (near-hypervelocity) impact (see **Methods** and **Fig. S11**). In this study, a hypersonic impact is defined as a collision event occurring at a relative speed ≥1700 m/s. For each test, a cylindrical composite sample was mounted in a custom two-piece 3D-printed axisymmetric target fixture (schematic in **Fig. 2a**). Samples were then tested under two conditions: (1) multiple aluminum 2017-T4 Ø1 mm simultaneously launched distributed particles (SLDPs) at an impact velocity of 2584.7 m/s (Mach 7.5) (**Fig. 2b**), and (2) a single aluminum 2017- T4 Ø4 mm sphere launched at 2900.8 m/s (Mach 8.45) (**Fig. 2e**). High-speed, high-contrast shadowgraphs qualitatively captured the entire impact event, including projectile flight, target impact, ejecta and debris cloud formation and expansion. The SLDP test produced only minor impact surface erosion with no visible cracking or fragmentation (**Figs. 2c**, **2d**, **S12** and **movie S3**), demonstrating the composite's high-hardness and superior damage tolerance under distributed loading. In contrast, the Ø4 mm sphere caused macroscopic fracture of the composite sample into several large pieces (**Figs. 2f-h**). In both tests, a fine, powder-like fast-moving front-face ejecta cloud (**Figs. 2c** and **2g**) emanated from the impact face, expanding in the direction opposite to the projectile's motion (i.e. uprange). For the Ø4 mm single-projectile impact, a distinct downrange debris cloud comprising larger fragments was expelled from the target back face (**Figs. 2g, 2h, S13,** and **movie S4**). In that case, the measured two-dimensional velocities of the leading edges of the front-face ejecta and downrange debris clouds were ~1122 m/s and ~487 m/s, respectively. Such phenomena, in which the ejecta cloud moves faster than the debris cloud, are relatively common in high-velocity impacts [55,56]. The fragmentation pattern, predominance of fine front-face ejecta, and relatively slower downrange debris are characteristic of impact-driven pulverization in quasi-brittle targets, typically occurring under hypervelocity conditions [57].

To get insights, fully atomistic reactive molecular dynamics (MD) simulations were carried out. An atomistic model of the diamond target, namely a 10×10×5 cubic-diamond supercell containing 4000 carbon atoms, impacted by an idealized projectile (**Fig. 2i**). The projectile, with a diameter of 20 Å, was initially placed at 15 Å above the target surface, corresponding to a center-of-mass standoff of 25 Å, and was then accelerated against the target with an initial velocity of 29 Å/ps along the -z direction, corresponding to ~Mach 8.45. An overview of the structural and bonding evolution of diamond during and after the projectile impact and the time-dependent evolution of the $sp^2$ and $sp^3$ hybridizations are shown (**Figs. 2j** and **S14**). We observed a decrease in $sp^3$ bond, and an increase in $sp^2$ bonds immediately after the impact, indicating the graphitization of diamond. Quantitatively, the $sp^3$ population of the pristine target falls to approximately 42% within 5 ps. In comparison, the $sp^2$ population reaches approximately 58% (**Fig. 2j**). Over the same interval, the bond-angle distribution shifts from 109.2° to 117.5°, the first-neighbor distance contracts from 1.57 Å to 1.48 Å, and the number of carbon atoms belonging to at least one six-membered ring reaches approximately 1300, close to 38% of the target atoms. Six-membered rings dominate, with 1136 atoms belonging to at least one hexagon against 219 and 708 atoms in five- and seven-membered rings, indicating a defective graphitic network rather than an amorphous phase (**Fig. S15**). Under extreme conditions, the tetrahedral network collapses into a more disordered, layered configuration. The similar transformation is also observed for a larger target (14×14×7 cubic-diamond supercell containing 10976 carbon atoms), indicating that the transformation is independent of the system size.

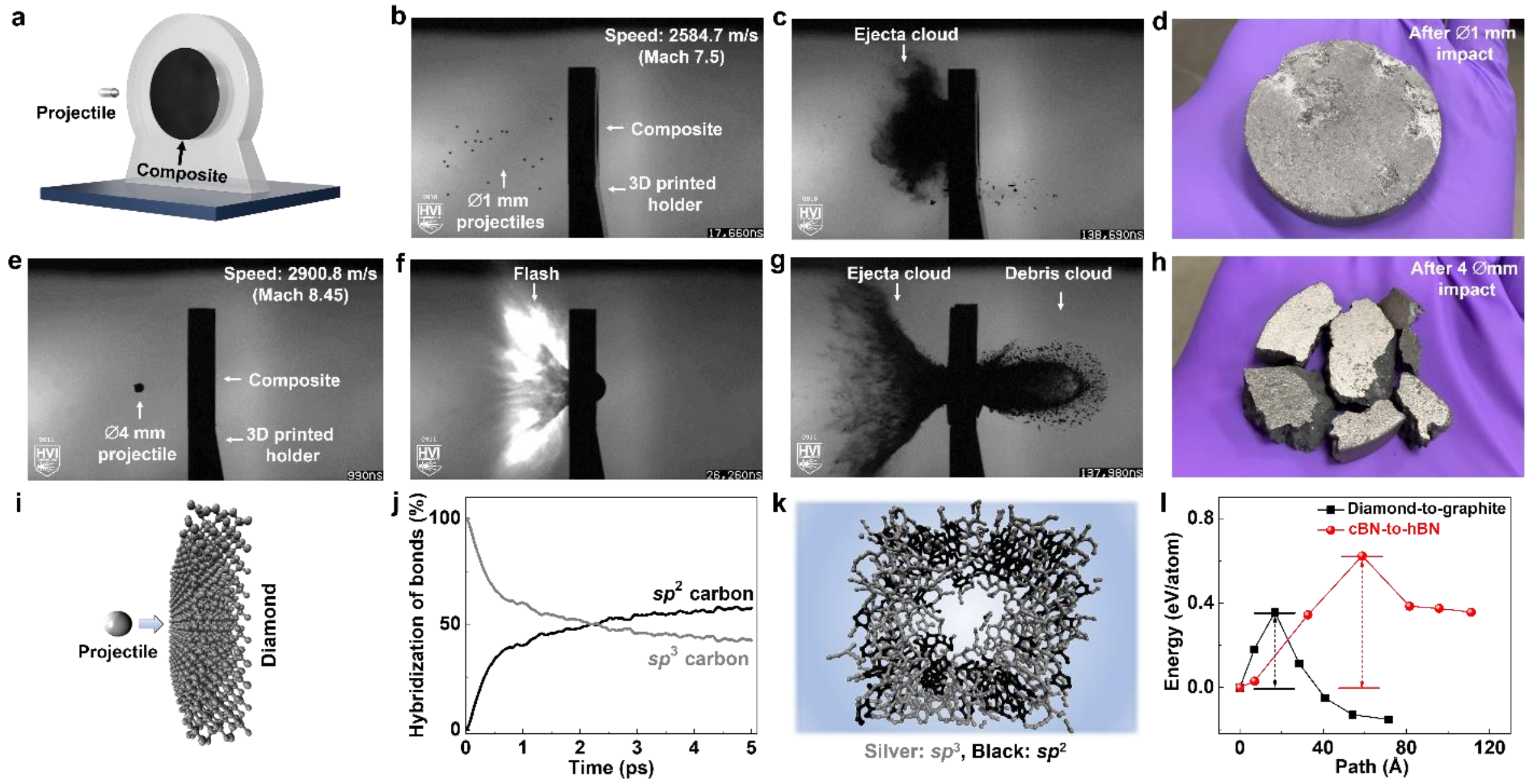

**Fig. 2. Hypersonic speed (near-hypervelocity) impact on the composite and its theoretical insights. (a)** Schematic of the impact test where a projectile hit the composite (black disk). The composite is placed onto a 3D printed holder, and spherical aluminum projectiles of Ø1 mm (speed ~2584.7 m/s, Mach 7.5) and Ø4 mm (speed ~2900.8 m/s, Mach 8.45) are impacted. The impact tests are conducted at ambient conditions. **(b), (c)** High-contrast shadowgraph of the impact event with ∅1 mm projectile. In the lower region, some ∅1 mm projectile missed the impact on the composite, and rather they impacted on the soft holder, creating debris. **(d)** The survived composite disk after the impact. **(e), (f), (g)** High-contrast shadowgraph of the impact experiment with ∅4 mm projectile, producing flashes, ejecta and debris clouds. **(h)** The ∅4 mm projectile impact completely shatters the composite disk. In the bottom right of each shadowgraph, the actual time scale of the event is provided. **(i)** Atomic structural model for the molecular dynamic simulations, showing a projectile impacting the diamond surface. The projectile was positioned 15 Å above the surface and shot against the target. **(j), (k)** The impact calculation shows a change in *sp*$^3$ (decreased) and *sp*$^2$ (increased) hybridization bonding of carbon and the formation of a hexagonal graphitic *sp*$^2$ carbon structure (black regions) due to impact. **(l)** The NEB energy barrier calculations for the phase transformation indicating that much higher energy is needed for the cBN-to-hBN phase transformation than diamond-to-graphite.

The phase transitions in carbon systems are governed not by a single universal energy threshold but by the combined effects of pressure, temperature, cooling rate, pressure-release rate, and crystallographic orientation [31,58,59]. Diamond graphitization, in particular, is orientation-dependent, with reported surface activation energies of 728 ± 50 and 1159 ± 75 kJ/mol for the (110) and (111) surfaces [32], and a graphitization resistance ordered (100) > (111) > (110) [60]. These simulations also indicate that impact-induced structural transformations in diamond proceed through complex atomistic pathways involving local bond rearrangement and intermediate metastable states. After the impact (**Figs. S14c** and **S14d**), the surface displays the emergence of interconnected graphitic layers and the formation of internal cavities during structural relaxation. The atomistic view shows the formation of hexagonal rings and graphitic ordering (**Fig. 2k**, and **movie S5**). The identification of graphitic domains was not based on visual inspection alone, since atoms with tetrahedral coordination (CN >3) were excluded (**Supplementary text)**, and six-membered carbon rings characteristic of graphitic structures were identified through a geometric-topological ring analysis, whose spatial distribution confirms the development of graphitic ordering. Regarding the phase transformation energy required, the Nudged Elastic Band (NEB) calculation shows that $sp^3$ diamond to $sp^2$ graphite forms an activation barrier of ~0.35 eV/atom (**Fig. 2l**), in close agreement with total-energy calculations of the concerted graphite-diamond transformation at zero applied pressure, which reported ~0.33 eV/atom [61], and in subsequent variable cell NEB work [62]. Substantially lower barriers are also reported, of ~0.08-0.25 eV/atom, correspond to nucleation-based pathways or to extreme hydrostatic pressure, where the barrier decreases as the lattice approaches instability [62,63]. In comparison, $sp^3$ cBN to $sp^2$ hBN transformation exhibits a higher energy barrier of ~0.62 eV/atom. These results reveal that both materials are kinetically protected metastable phases, however cBN exhibits a higher resistance (requiring more energy) to hBN phase transformation, compared to diamond-to-graphite. Taken together, the decrease in the number of $sp^3$-type bonds and the increase in $sp^2$-like ones, the emergence of hexagonal graphitic rings, the formation of interconnected graphitic layers after structural relaxation, and the relatively low NEB barrier establishes the impact-induced graphitization process.

*Characterizations of the composite after ∅4 mm projectile impact test*

Post-impact, as the composite disk was completely broken (∅4 mm projectile impact case); the fragments (**Fig. 2h**) were retrieved from the target tank for characterization. XRD shows a highly intense Graphite G (002) peak along with a much lower intensity D (111) and cBN (111) peaks (**Fig. 3a**). In XRD, the graphite peak is strong, but the formation of hBN phase was not detectable, however, locally transformed hBN phase may exist in the composite which does not constitute the major phase. For comparison, XRD and Raman spectroscopy of ∅1 mm SLDP impact survived composite, also showing partial graphitization (**Fig. S16**). For the ∅4 mm projectile impact case, pole figure map of the (002) Bragg peak of the fractured composite shows oriented graphite basal planes (**Fig. 3b**), suggesting a certain preferred texture in the diamond transformed graphite. The pole figure map (compared with pristine composite) of the D (111) and G (002), shows some randomness of diamond particles but a more oriented graphite formed after the impact (**Fig. S17**). XRM imaging shows a high volume of graphite (black) throughout the composite and low volumes of diamond having smaller grain sizes (red) (**Fig. 3c** and **movie S6**). Hypersonic impacts most likely effectively reduced the Co-cluster (particle) sizes to less than the resolution limit of XRM, thus it is difficult to observe many Co clusters in the XRM image. EPMA SEI shows the topography of the fractured composite sample after the impact (**Fig. 3d**) with small protrusions of diamond particles in the center and a flatter morphology of the graphite flakes. The phase map combined with the WDS elemental maps show carbon surrounded by cBN-Co matrix (**Fig. 3e**). The core-level XPS shows the possible bonding and slight peak shift (~0.12 eV) of the most intense peak in C 1s to lower binding energy, indicating the formation of $sp^2$ phase (**Fig. S18**). The large-area FESEM morphology shows almost flat layered-sheet-like surface morphology throughout the surface (**Fig. 3f**), very different from the pristine rougher surface with the larger diamond particles. One can clearly see a graphitic layered sheet (almost similar grain size of diamond), and its EDS map further confirming that this graphite (carbon) region is surrounded by cBN (**Figs. 3g** and **3h**). Images from several other regions (both from the impact surface and the fractured lateral edges) also show formation of layered graphite (**Fig. S19**). The non-carbon regions show the presence of both cBN and Co (**Figs. 3h** and **S20**). Raman spectroscopy and map on graphitic layers show the characteristic G-band peak at ~1579.70 $cm^{-1}$ along with a defect D-band peak at ~1345.06 $cm^{-1}$ (**Fig. 3i** and **Fig. S21**). Further in the cBN-Co region, Raman spectroscopy shows TO and LO mode peaks, suggesting the presence of cBN particles [27] (without any signature of hBN peak in the Raman spectra) (**Fig. 3j**). Considering the formation of graphitic layers, the composite now shows more than six-fold increase in electrical conductivity (resistivity ~36 mΩ-cm), and retains its ferromagnetic behavior at room temperature (**Fig. S22**), however with slightly reduced moment due to the reduction in Co fraction, as seen from the WDS elemental map (**Fig. S23**), due to possible removal of Co during the impact event.

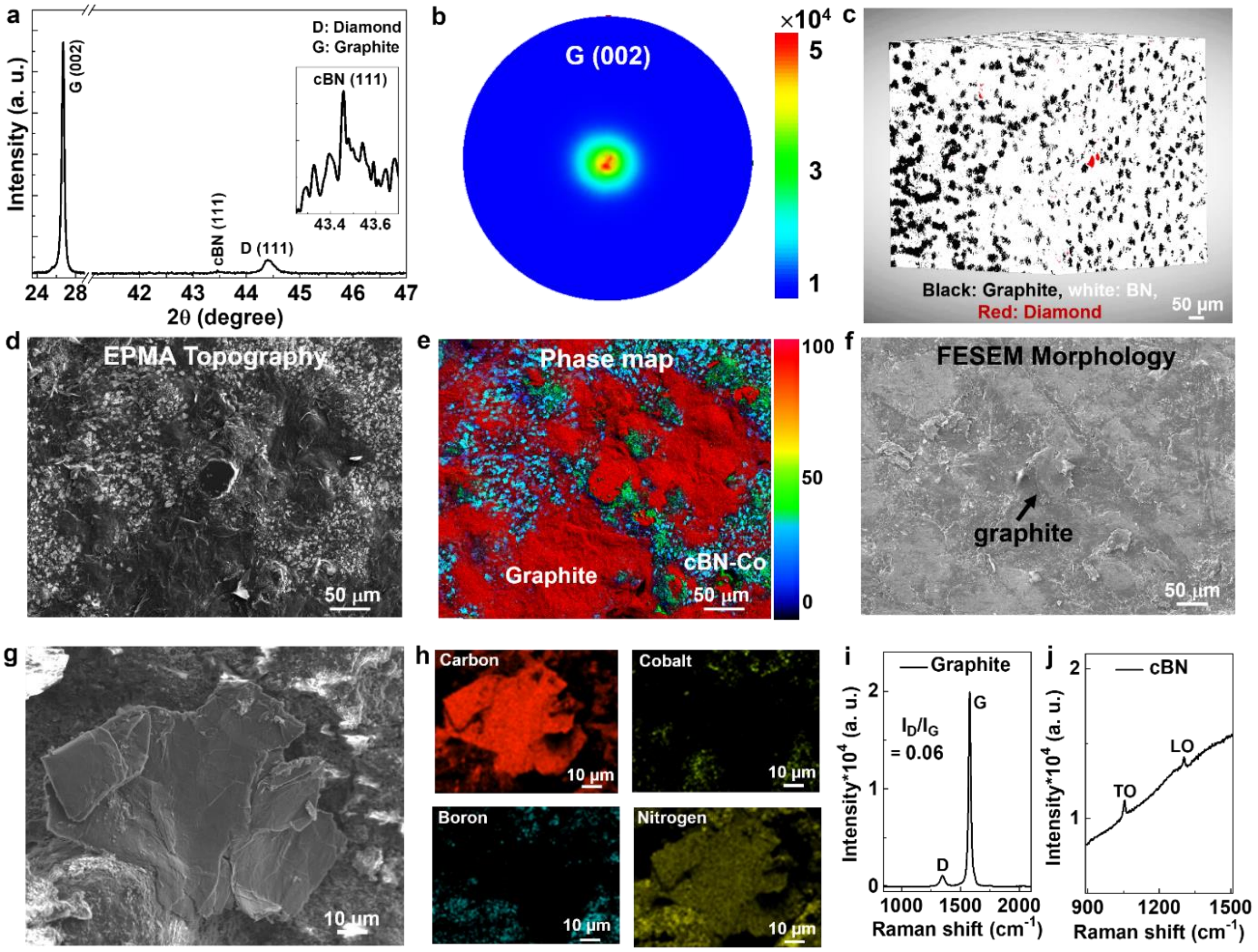

**Fig. 3. Characterizations of the composite after the ∅4 mm projectile impact. (a)** XRD shows the highly intense graphite G (002) peak and lower intense diamond (D) and cBN (111) peaks. Inset shows the zoomed-in region of the cBN peak. **(b)** Pole figure map of G (002) shows oriented graphite throughout the bulk. **(c)** XRM imaging shows the formation of a high volume of graphitic (black) regions throughout the bulk with a tiny presence of diamond (red). **(d)** EPMA secondary electron image showing the topography, with a protrusion of smaller diamond-like features in the center, and topography of flatter, larger graphite flakes, surrounded by brighter cBN-Co matrix.

**(e)** Phase map resulting from the combination of WDS element maps shows diamond-graphite (red), cBN (green), and light blue (Co). **(f)** Large-area FESEM image shows flat surface with layered morphology. **(g), (h)** Image of a graphite layer (almost similar sizes of diamond particle) embedded within the cBN-Co matrix. **(i)** Raman spectroscopy shows characteristic G-band (~1579.07 $cm^{-1}$), and disordered D-band (~1345.6 $cm^{-1}$) peaks of graphite. **(j)** Raman spectroscopy from non-graphitic region shows TO (~1053.64 $cm^{-1}$) and LO (~1304.06 $cm^{-1}$) phonon modes, characteristics of cBN.

### *Atomic-scale insights of the diamond-to-graphite transformation*

An atomistic understanding of graphitization of diamond, during impact is important. We investigated the partially graphitized diamond particles to reveal the crystallographic relationship of the transformed graphite layers starting from diamond. In many regions, the FESEM image of diamond particles in the after-impact sample shows the formation of layered sheet-like features around the edges of diamond particles (**Figs. 4a**, **S24**, and **S25a**), with the surrounding regions still composed of cBN. EPMA SEI also shows particles that resemble diamond in the original composite in size and shape; however, a closer look at the lower region of the image shows the formation of layered sheet-like features indicating the formation of graphite (**Figs. S25b** and **S25c**). The cathodoluminescence (CL) image from a random region shows two distinct emitted intensity profiles from the diamond grain. The intensely red-colored region corresponds to diamond, whereas graphite shows almost no luminescence (**Fig. 4b**). In contrast, the intermediate region showing light-blue towards green indicates the partial transformation of graphite-to-diamond. We also observed the X-ray $k_\alpha$ peak shift from EPMA of carbon for diamond, diamond partially transformed to graphite, and diamond fully transformed graphite, showing distinct peak shifts that are consistent with the graphite EPMA standard reference (**Fig. 4c**). The peak shift reflects the change in the bonding and re-hybridization of carbon during the transformation of diamond-to-graphite [64].

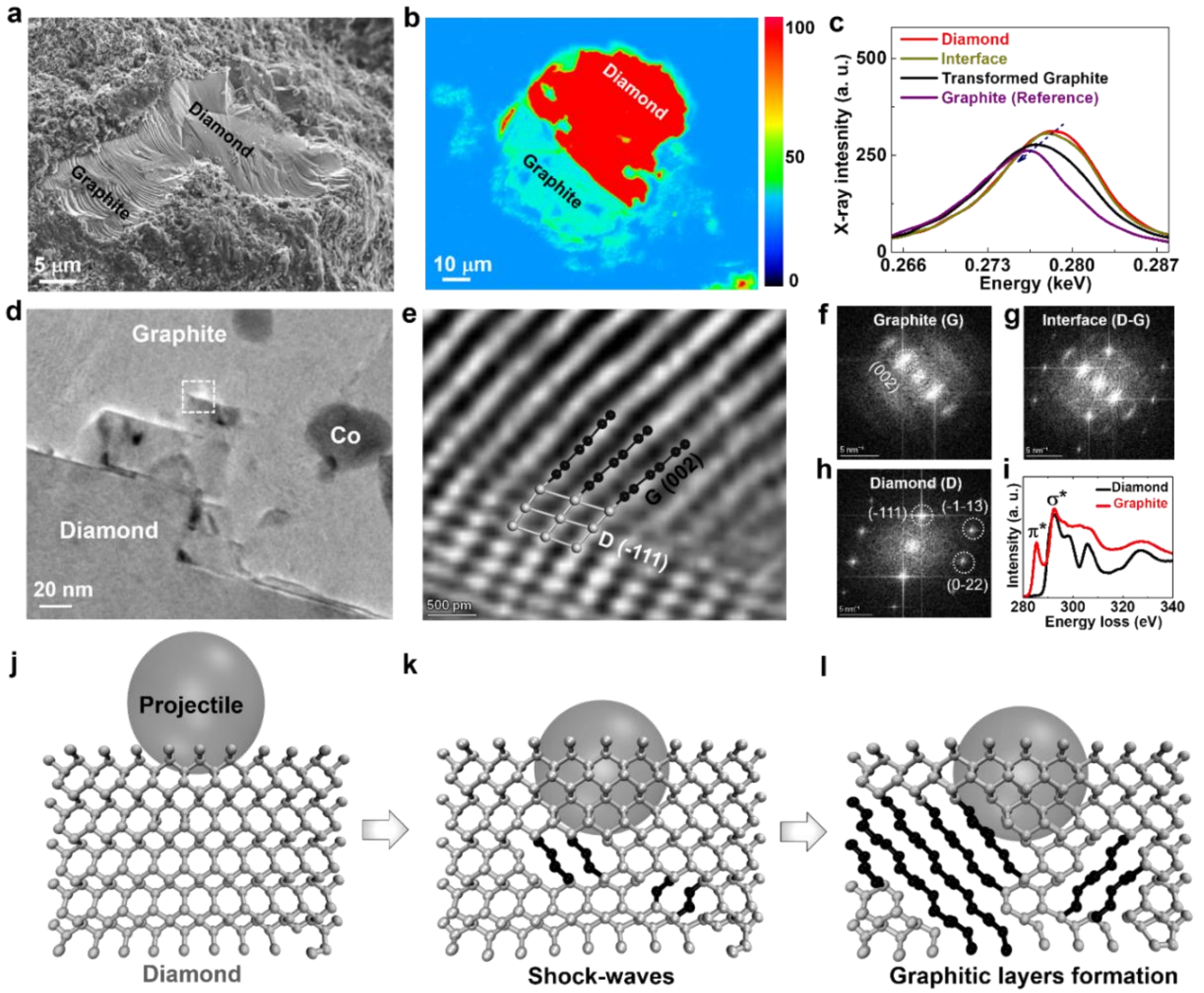

**Fig. 4. Microstructural insights of the diamond-to-graphite transformation. (a)** FESEM showing the layered graphite sheet-like textures within diamond. **(b)** Cathodoluminescence image shows the emitted intensity from a diamond grain in the visible wavelength range. The red color area represents diamond, the light-blue towards green represents the partially transformed graphite, while the darker light blue is the fully transformed graphite. **(c)** X-ray $k_{\alpha}$ peak shifts of carbon from the regions of diamond, partially transformed graphite (near interface), fully transformed graphite, and EPMA standard graphite as a reference. The peak shift attributes the transition from diamond-to-graphite. **(d)** Bright-field TEM image shows the interface of diamond and graphite. **(e)** HRTEM image (white box region in **d**) shows the atomic arrangements and interface between diamond and graphite. Atomic structure is shown for clarity with diamond (silver) and graphite (black). **(f), (g), (h)** FFTs patterns from diamond, graphite and the interface showing crystallographic orientation. **(i)** STEM EELS reveal $sp^2$ and $sp^3$ characteristics of diamond and graphite, respectively. **(j), (k)**

MD simulations using the ReaxFF show that diamond structure breaks the bonds due to shear (black regions). **(l)** Fully graphitic layer's formation (black colored for visual representations) due to hypersonic impact, creating diamond-graphite interfaces.

We performed high-resolution transmission electron microscopy (HRTEM) to capture the diamond-to-graphite phase transformation, which can provide insight into the nature of the phase transformation of diamond particles under impact. The low-magnification bright field (BF) TEM image acquired from cross-sectional focused ion beam (FIB) lamella shows distinct regions of diamond (D), graphite (G) and Co nano-domains, confirmed by the HAADF-STEM image (**Figs. 4d** and **S26**). The concurrent phase formation of diamond and graphite (the region of interest) was investigated by precession 4D-STEM and diffraction patterns from the respective regions showing [211] and [110] orientations of diamond and graphite lattice (**Fig. S27**). High-magnification HRTEM shows atomic-scale images with one-to-one correspondence between diamond and graphite lattice (**Fig. 4e**). It shows the D (-111) to G (002) atomic orientation relationship at the interface (in the figure we have indicated the structure for clarity). The D-G interface might look semi-coherent locally, but overall it is not coherent as dislocations can be seen (bottom right of **Fig. 4e**). The lattice mismatch is also noticeable and the interface is bending, so the semi-coherency is changing along the interface. The interlayer spacing of graphite varies across the region due to local bending and it is found to be ~3.4±0.1 Å, the (002) *d*-spacing of bulk graphite phase. The interlayer spacing remains consistent near the diamond-graphite interface, however increasing deviations are observed away from the interface [30].

The interlayer *d*-spacing between diamond layers is ~2.09±0.05 Å, suggesting that the lattice is strained (as bulk diamond $d_{(-111)}$ = 2.05 Å). The corresponding Fast Fourier Transformations (FFTs) from the diamond, graphite and the interface region confirm the crystallographic relation of D (-111) and G (002) planes (**Figs. 4f-h**). The STEM EELS show the characteristic carbon spectra with both $\pi^*$ (285.4 eV) and $\sigma^*$ (292.5 eV) peaks of $sp^2$ bonded graphite, whereas only the $\sigma^*$ peak of $sp^3$ bonded diamond (**Fig. 4i**). While graphite forms from the diamond lattice under impact, interestingly, we also observe misorientations between the graphite layers, as confirmed by the diffraction patterns obtained from 4D-STEM of the graphite nano-region with random mis-match angles of 7° and 32° with some defects in the form of dislocations and stacking faults (**Fig.**

**S28**). Rapid impact can induce significant orientation changes between the graphite layers disrupting the natural AB-ordered stacking. Under shock produced during impact, basal planes could rotate or buckle as dislocations and lattice defects proliferate (producing non-coherent interfaces). Further, the impact can also create localized regions of twisting and tilting between adjacent sheets, producing misalignments, as observed here in certain regions. Further, in relatively few locations (compared to the large-scale phase change that has happened to diamond), we also observed localized phase conversion of cBN having $d_{(111)}$ of ~2.05 Å to hBN with $d_{(002)}$ of ~3.5±0.2 Å (**Fig. S29**), though XRD and FESEM (**Figs. 3a** and **S20**) showed that predominantly cBN phase was retained after impact. The cubic lattice of Co appeared to remain intact even after the impact (**Fig. S30**).

The impact-induced diamond-to-graphite phase transformation should be an extremely rapid phase change occurring over microseconds. During the impact event, the energy absorption could take place in multiple ways. First, fragmentation and formation of large fractured surfaces could dissipate energy via free surface energy creation; this does not appear to be significant, as only a few large target fragments were produced during impact with a single ∅4 mm projectile and the composite sample did not fracture when impacted with ∅1 mm SLDPs. Second, energy absorption could happen via strain-induced massive plastic deformation, which also does not seem to be the case here. Third, significant energy absorption could result from large-scale diamond-to-graphite phase change; we believe this is the primary mechanism occurring in these impact experiments. The fact that the transformation is extremely rapid, we hypothesize that for this particular composite system nonequilibrium pathways driven by shock-waves produced during the impact lead to massive reorganization of the carbon lattice, with diamond transforming into graphite. However, being a complex system, there could still be a hybrid effect of both shock-induced graphitization and temperature-induced graphitization as Co can catalyze the graphitization process within microseconds as well. We investigated the early stage graphitization mechanism using MD simulations with the system modeled using a reactive force-field (ReaxFF). Such a high-speed impact generates intense shock-waves that induce instantaneous pressure spikes followed by rapid release, shear, and localized thermal spikes, concurrently occurring in microseconds (whereas traditional graphitization relies on sustained high temperatures and diffusion processes). This shock generates an intense compressive zone beneath the point of contact, followed by the rapid propagation of elastic waves of shear stresses throughout the diamond lattice (**Figs. 4j-l** and **movie S7**). This shear deformation combined with localized thermal spikes forces the $sp^3$-bonded tetrahedral framework to collapse into $sp^2$-bonded graphitic layers before the system can equilibrate. The highly stressed regions act as nucleation sites for graphitic layers, as the local collapse of the tetrahedral $sp^3$ framework facilitates the reorganization of carbon atoms into planar $sp^2$ configurations. As time progresses, the stress field becomes anisotropic, accumulated along the main crystallographic directions favoring structural rearrangement to produce diamond-to-graphite phase transformation.

## Conclusions

We have found a way to stabilize diamond during spark plasma sintering using a unique combination of a matrix material (cBN) and stabilizer (Co), enabling us to create diamond-based hard composites, which are otherwise impossible to produce via sintering. This composite was subjected to hypersonic speed impacts using metal projectiles. During the impact, energy transfer, and fracture event, diamond grains embedded in the composite undergo instantaneous phase transformation to graphite. Atomic-scale structural analyses and atomistic simulations provide insights into the diamond-to-graphite phase transitions and formation of diamond-graphite interfaces during the impact. This impact-induced phase transformation is uniquely important because it probes matter under extreme conditions, revealing fundamental insights into diamond's bonding dynamics, structural instability, and rapid phase kinetics, producing different architectures. Our results provide a unique example of large-scale diamond-to-graphite transition under extreme conditions of impact and shock. Finally, our findings would also be important in designing novel diamond-based composite materials, suitable for extreme environments.

## Methods

### *Spark plasma sintering of diamond-cubic BN-Co composite*

We performed spark plasma sintering (SPS) by using commercially available yellow color diamond powders [12,65] (particle size: 50~100 μm, CAS# 7782-40-3, Advalue Technology, Tucson, AZ, USA), gray color cubic BN powders (particle size: 1 μm, SKU: PO6901, MSE Supplies, Tucson, AZ, USA), and black color cobalt powders (particle size: 1.6 μm, 99.5% trace metals basis, CAS# 7440-48-4, BeanTown Chemical, USA). SPS was conducted using an SPS 25-10 system (Thermal Technology LLC, California, USA) at Texas A&M University, USA. A constant uniaxial pressure of 90 MPa was applied (maximum achievable), with a heating rate of 50 °C/min, reaching a maximum temperature of 1400 °C. For SPS, equal wt% diamond, cubic BN and cobalt powders were mixed and ground for 15 min in a mortar pestle. The mixed powder was wrapped in graphite foil and placed inside the graphite die and placed into the SPS chamber under an initial pressure of 5 MPa. Graphite foil serves as a barrier to prevent reactions between the sample and die, while also enhancing thermal distribution and insulation. The chamber was pre-evacuated to ~$2\times10^{-5}$ Torr for about 30 minutes, followed by sintering for 60 minutes under ultra-high purity argon gas (5N, ~99.999%). During SPS, a direct current passes through the material, generating the Joule heating process. The temperature during the sintering was monitored using an optical pyrometer (Raytek, Berlin, Germany, model D-13127). After sintering, the pressure was released gradually at ~5 MPa/min, while the sample was cooled at ~100 °C/min. We used large size diamond particles as smaller particles (≤ 0.5 μm) with a higher surface area-to-volume ratio makes them more susceptible to surface graphitization because of high surface energy, making more graphitic phases. In addition, while SPS is often associated with rapid sintering and short dwell times, in the present study, the 1 h holding time was used to enhance densification and microstructural homogeneity as it's a multiphasic composite, and also, we used a relatively low sintering temperature. Regarding the particle size dependence, for one diamond-cBN-Co composite, we used smaller particle sizes diamond powders (Batch: 3182025KP, MSE Supplies, Tucson, AZ, USA), however, same particle size cBN and Co powders and similar SPS conditions. For two other composites (Diamond-cBN, and diamond-Co), we used larger size diamond, same Co powders and similar SPS conditions, however used larger grain sizes yellow color cBN powders (particle size: 36-54 μm, Batch: 01723C4, MSE Supplies, Tucson, AZ, USA).

***Structural characterizations (XRD, Pole figure, XPS, Raman spectroscopy, FESEM)***

X-ray diffraction (XRD) was carried out using a Rigaku SmartLab thin-film diffractometer (Tokyo, Japan), operating at 40 kV and 40 mA with a monochromatic Cu $K_\alpha$ radiation source ($\lambda$ = 1.5406 Å), and a scan rate of 0.05°/min (short range scan) and 0.2°/min (wider range scan). For the pole figure map, we used the same XRD instrument in which we performed Chi/phi scans by fixing the detector at the 2θ positions of the respective phases. We performed X-ray photoelectron spectroscopy (XPS) by using a PHI Quantera SXM scanning X-ray microprobe equipped with a monochromatic Al $K_\alpha$ source (1486.6 eV). High-resolution core-level elemental spectra were obtained at a pass energy of 26 eV. We used MultiPak V6.1A software to fit the spectra. Raman spectroscopy was conducted using a Renishaw inVia confocal microscope equipped with a 532 nm excitation laser. Surface morphology and the energy dispersive X-ray spectrometry (EDS) mappings were analyzed by field emission scanning electron microscopy (FESEM) using a JEOL JSM IT800 SHL FEG system. For FESEM imaging the energy used was 3 kV whereas for high-resolution EDS maps the energy used was 20 kV.

***X-ray microscopy (XRM)***

X-ray microscopy (XRM) images (voxel size ~1.5 µm) were acquired using a Sigray Eclipse 900 XRM system equipped with an ultrahigh-resolution nanofocus X-ray source operating at 60 kV and a 27-MP large-field-of-view flat-panel detector. Image processing was subsequently performed with 3D Dragonfly software, and a cropped region is presented for clarity. Phase segmentation of diamond and graphite was conducted using the machine learning (ML) algorithm integrated within the software.

***Quantitative electron probe micro-analysis (EPMA)***

Quantitative EPMA data acquisition was carried out on a Jeol JXA 8530F Hyperprobe, equipped with a Schottky field emission gun and five Wavelength Dispersive Spectrometers (WDS). Both samples (before and after impact) were coated with amorphous carbon to eliminate the eventual discharge, assuring ground conductivity. The carbon film thickness was 25 nm, similar to the carbon film deposited on the reference standards used in the quantitative analysis (diamond, BN, and Cobalt metal). Analytical conditions employed for analysis were an accelerating voltage of 15 kV, beam current of 50 nA, and spot beam size (~350 nm diameter). The PAP method was employed for quantification [66]. The experimental parameter details are presented in Supporting Information **Table S1**. Detector conditions for each element are shown in **Table S2**. Quantitative Wavelength Dispersive Spectrometry (WDS) elemental maps were acquired at 15 kV accelerating voltage and 50 nA beam current, using stage mode with 200 ms dwell time. Different elements were mapped by recording the relative intensity of their characteristic X-ray line (**Table S1**). The same standards used for quantitative analysis were employed for the quantification of element maps. Deadtime correction was applied for each element map. The map resolution is 550 × 470 pixels, with 1 pixel = 1.2 μm, or the size 660 μm × 564 μm. WDS qualitative analysis (WDS scans) employed the same accelerating voltage and beam current as in the quantitative analysis. The scan interval was between 0.264 to 0.290 keV, at a step of 0.000066 keV. The peaks were smoothed using the Savitzky-Golay method ('auto smoothing' mode).

### *Electrical transport and magnetization*

The temperature-dependent AC resistivity between 2-300 K was measured using the Quantum Design Physical Property Measurement System (PPMS) employing the standard four-point probe method with silver point contacts. A constant current of 2 mA was applied, and the corresponding voltage was recorded while the temperature was swept at a rate of 1 K/min. The magnetoresistance (MR) was performed with the electrical current applied in-plane (*ab*-plane) and the magnetic field applied oriented within the *ab*-plane over a field range of -5 to 5 T.

Field-dependent magnetization (M-H) was measured at 300 K using the DC measurement mode of a Quantum Design Magnetic Property Measurement System (MPMS 3) equipped with a 7 T superconducting magnet. The sample was mounted in a standard plastic straw holder, and the M-H loops were collected between -5 and 5 T at a field sweep rate of 200 Oe/sec using the auto-centering option.

### *Hypersonic speed (near-hypervelocity) impact test*

The impact experiment was performed at Texas A&M University (TAMU) Hypervelocity Impact Laboratory (HVIL) using a two-stage light-gas gun (2SLGG) and aeroballistic range. The gun is capable of launching large numbers of 0.1-2.0 mm variable-shaped particle clusters ("simultaneously launched distributed particles" or SLDPs) or wide variety of 1.6-10 mm diameter single projectiles of various geometries (spherical, cylindrical, ogival, etc.) at velocities of 2-8 km/s. The 2SLGG is capable of accelerating projectiles at hypervelocity (typically ≥3 km/s) by compressing a light working gas (e.g., hydrogen) to high pressure using a gunpowder charge and a consumable polymer piston. In the first stage, the combustion gases from the gunpowder ignition propels the polymer piston down an enclosed pressurized tube (i.e., "pump tube") containing the light working gas. Once the light gas is compressed to a critical pressure (typically 45-63 MPa), a diaphragm at the downrange end of the pump tube ruptures, releasing the high-pressure gas that accelerates the projectile package down a launch tube to the desired velocity. The light gas compression in the pump tube constitutes the second and final stage necessary to accelerate the projectile. A more complete overview of the TAMU HVIL 2SLGG facility, including the operational details and diagnostic capabilities are provided (**Fig. S31**) [67].

The cylindrical composite disks (∅40 mm, and ~7.4 mm and ~6.6 mm thick, respectively) were mounted in custom two-piece 3D-printed (polylactide) axisymmetric target fixtures. The fixture contained a central circular aperture such that the impact- and back-faces of the sample were clamped between the two pieces using four evenly spaced fasteners. This target assembly was secured to a mounting fixture in the target tank, with the sample centered about the projectile flight axis. The disks were then impacted with several Ø1 mm (total mass of 0.049 g) at 2584.7 m/s and a single Ø4 mm 2017-T4 aluminum sphere (mass of 0.091 g) at 2900.8 m/s. The projectile's velocity was measured using a dual-laser-curtain velocimetry system, with an accuracy of ±2 m/s. A total of 128 shadowgraphs of each impact event were recorded using a Shimadzu HPV-X2 operating at 751,880 frames per second with an interframe exposure time of 200 ns. The open-source Tracker motion-tracking software (version 6.3.2) was used to determine the 2D velocity of the expanding leading edges of ejecta and debris clouds from the high-speed shadowgraphs.

### *Transmission electron microscopy (HRTEM)*

A double aberration corrected Themis Z (ThermoFisher Scientific) TEM equipped with a Super-X energy dispersive X-ray detector (EDX) and a Gatan GIF Continum 970 HighRes camera was used for atomic scale characterization at an accelerating voltage of 300 kV. A cross sectional TEM lamella was prepared by focused ion beam using Helios dual beam (ThermoFisher Scientific) FIB. High resolution imaging was performed in both transmission (TEM) and scanning (STEM) mode.

STEM and TEM images were acquired using a high-angle annular dark field (HAADF) and annular bright field (ABF) detector and Ceta 16 M camera. Electron energy loss spectra (EELS) in STEM were collected with 2.5 mm aperture at an energy dispersion of 0.1 eV per channel and 1.1 eV energy resolution. 4D-STEM measurements for phase and orientation mapping were carried out using NanoMegas ASTAR system with a precession electron diffraction (PED). The microscope operated in micro-probe STEM mode with a 0.5 mrad convergence angle and a camera length of 245 mm. The precession angle was set to 0.6° and diffraction patterns were collected using a pixelated Dectris Quadro Detector.

### *Theoretical calculations*

### *Energy barrier calculations*

The energy interconversion barriers were estimated using Nudged Elastic Band (NEB) simulations [68]. We employed the Self-Consistent Charge Density Functional Tight-Binding (SCC-DFTB) method, as implemented in the DFTB+ code [69], using the Slater-Koster matsci-0-3 parameter set. Diffusion barriers were computed using the NEB method as implemented in the Atomic Simulation Environment (ASE) framework [70]. The diamond (or cBN) and graphite (or hBN) structures were used as initial and final configurations, respectively, with five interpolated images generated along the minimum energy pathway (MEP). The Improved Dimer Projection Path (IDPP) algorithm [71] was applied to refine the initial interpolation, ensuring a more accurate representation of the transition pathway and saddle-point region. Geometry optimizations for all images were carried out using the BFGS algorithm until the maximum residual force on any atom was below 0.05 eV/Å, yielding well-converged transition-state configurations.

### *Molecular dynamics (MD) simulations using a reactive force-field (ReaxFF)*

To investigate the insights about the projectile impacting on a diamond surface, fully atomistic reactive molecular dynamics (MD) simulations were carried out using the LAMMPS code [72]. The systems were modeled using the reactive force field ReaxFF [73], which accurately describes bond formation and dissociation, bond-order variations, and charge redistribution. Charge equilibration was performed at every integration step using the qeq/reaxff algorithm with a convergence tolerance of $10^{-6}$, ensuring consistency in energy, forces, and atomic charges, even under highly nonlinear conditions typical of hypersonic regimes.

The simulation cell was a cubic box with dimensions of 200 Å, with periodic boundary conditions along all directions. The diamond target was created from a 10×10×5 supercell, containing 4000 atoms. The effect of the system size was assessed by repeating the impact simulation on a larger target created from a 14×14×7 supercell, containing 10976 atoms, under otherwise identical conditions. The idealized projectile center of mass was positioned 25 Å above the target surface and assigned an initial velocity of 29 Å/ps along the -*z*-direction, corresponding to approximately Mach 8.45. Its mass was set equal to that of the substrate to maximize momentum transfer and impact intensity.

The equations of motion were integrated using the velocity-Verlet algorithm [74] with a timestep of $1.0\times10^{-4}$ ps during thermal equilibration and $5.0\times10^{-5}$ ps during the impact event. This choice ensured numerical stability in the presence of large forces and rapid charge redistribution inherent to the ReaxFF potential. To prevent computational errors in bond detection under extreme compression, the ReaxFF implementation employed extended safety buffers, thereby enhancing numerical robustness without altering the system's physical behavior.

The simulation protocol began with an energy minimization using a convergence criterion of $10^{-6}$ for both energy and forces to remove any residual atomic overlaps. Subsequently, a 10 ps equilibration MD run was performed in an NVT ensemble, maintaining the system at 300 K using a Nosé-Hoover thermostat [75,76]. After equilibration, the simulation was changed to an NVE ensemble to ensure energy conservation during the impact event. The projectile interacted with the targets via a purely repulsive Weeks-Chandler-Andersen (WCA) potential [77,78], designed to eliminate any attractive components that could lead to artificial adhesion or unintended chemical reactions. The WCA parameters were set to $\sigma$ = 10 Å and $\varepsilon$ = 5 eV, with a truncation at $r_c = 2^{1/6}\sigma$ and an energy shift such that $V(r_c) = 0$, ensuring a smooth potential cutoff.

It should be noted that the present molecular dynamics simulations were designed to investigate the local atomistic mechanisms of impact-induced structural transformation in diamond, namely bond rearrangement and graphitization, rather than the macroscopic impact behavior. Accordingly, the analysis focuses on the structural modifications occurring in the vicinity of the impact region.

**Acknowledgements**

P. M. A. acknowledges the support from Air Force Office of Scientific Research (AFOSR) under award number FA9550-24-1-0301. The authors gratefully acknowledge the support of the Texas A&M University (TAMU) Hypervelocity Impact Laboratory (HVIL) for enabling the impact experiments. R. S. and C. K. are grateful to the Karlsruhe Nano Micro Facility (KNMFi) at KIT for providing TEM facilities. L. Z. D. acknowledges the funding support provided by the Robert A. Welch Foundation (00730-5021-H0452-B0001-G0512489). S. K., C. W. C. and L. Z. D acknowledges support from the U.S. Air Force Office of Scientific Research Grants FA9550-15-1-0236 and FA9550-20-1-0068; the T. L. L. Temple Foundation; the John J and Rebecca Moores Endowment; and the State of Texas through the Texas Center for Superconductivity at the University of Houston. T. F., L. L., Y. Z., and S. Y. acknowledges financial support from the Natural Sciences and Engineering Research Council of Canada (NSERC) and the Canada Foundation for Innovation (CFI). G. C. acknowledges the use of the EPMA facility at the Department of Earth Science, Rice University. Marcelo L. Pereira Junior acknowledges financial support from the Federal District Research Support Foundation (FAPDF) under grant 00193-00001807/2023-16, from the National Council for Scientific and Technological Development (CNPq) under grants 444921/2024-9 and 308222/2025-3, and from the Coordination for the Improvement of Higher Education Personnel (CAPES) under grant 88887.005164/2024-00. Raphael B. de Oliveira thanks CNPq for support under processes 151043/2024-8 and 200257/2025-0. Guilherme S. L. Fabris acknowledge support from the São Paulo Research Foundation (FAPESP) through fellowship 2024/03413-9. Douglas S. Galvão acknowledges support from FAPESP under grant 2025/27044-5 and from the Research, Innovation, and Dissemination Centers (CEPID) program under grant 2013/08293-7. It is also acknowledged for support from INEO/CNPq and FAPESP grant 2025/27044-5. We also thank the Coaraci Supercomputer Center for providing computational resources under process 2019/17874-0. A. B. thanks Texas A&M user facility and A. Bandyopadhyay for the SPS. P. M. A. and A. B. would like to thank Ramon Torrecillas, Jacob Elkins, and Atin Pramanik for their technical helps. X-ray facilities managed by the Shared Equipment Authority of Rice University.

## AUTHOR DECLARATIONS

### Competing interests

The authors declare that they have no known competing financial interests or personal relationships that could have appeared to influence the work reported in this paper.

### Author contributions

A. B. and P. M. A. conceptualized the work. A. B., J. M., S. C., S. Y., L. L., Y. Z., T. F., and J. L. performed the materials synthesis and characterizations. A. M., A. B, P. M. A, and T. L. Jr. performed the impact tests. R. S. and C. K. performed the HAADF-STEM. M. L. P. J., R. P. O., G. S. L. F., and D. S. G. performed the theoretical calculations. G. C. performed the EPMA. S. K., C. W. C., and L. D. performed the electrical transport measurements. A. B., R. V., and P. M. A. analyzed the data. A. B., and P. M. A. wrote the paper with the help of all the authors. All the authors discussed the results and contributed to the manuscript preparation.

### Data availability

The data that support the findings of this study are available from the corresponding author upon reasonable request.

### Appendix A. Supplementary material

Supplementary material contains various experimental results (HRTEM, XRD, XRM, FESEM, Raman spectroscopy, EMPA, electrical and magnetic transport measurement, and projectile impact testing) and theoretical calculations.

### Appendix B. Supplementary videos

Supplementary videos to this article can be found online.

## References

[1] Q. Huang, D. Yu, B. Xu, W. Hu, Y. Ma, Y. Wang, Z. Zhao, B. Wen, J. He, Z. Liu, Y. Tian, Nature 510 (2014) 250–253.
[2] T. Irifune, A. Kurio, S. Sakamoto, T. Inoue, H. Sumiya, Nature 421 (2003) 599–600.
[3] J.E. Graebner, S. Jin, G.W. Kammlott, J.A. Herb, C.F. Gardinier, Nature 359 (1992) 401–403.
[4] X. Guo, M. Xie, A. Addhya, A. Linder, U. Zvi, S. Wang, X. Yu, T.D. Deshmukh, Y. Liu, I.N. Hammock, Z. Li, C.T. DeVault, A. Butcher, A.P. Esser-Kahn, D.D. Awschalom, N. Delegan, P.C. Maurer, F.J. Heremans, A.A. High, Nat. Commun. 15 (2024) 8788.
[5] J.E. Field, Rep. Prog. Phys. 75 (2012) 126505.
[6] A. Nie, Z. Zhao, B. Xu, Y. Tian, Nat. Mater. 24 (2025) 1172–1185.
[7] J. Jing, F. Sun, Z. Wang, L. Ma, Y. Luo, Z. Du, T. Zhang, Y. Wang, F. Xu, T. Zhang, C. Chen, X. Ma, Y. He, Y. Zhu, H. Sun, X. Wang, Y. Zhou, J.K.H. Tsoi, J. Wrachtrup, N. Wong, C. Li, D.-K. Ki, Q. Wang, K.H. Li, Y. Lin, Z. Chu, Nature 636 (2024) 627–634.
[8] J. Tu, J. Li, Y. Wang, Y. Zhao, J. Liu, J. Wei, L. Chen, J. Zhang, Y. Lu, C. Li, Nat. Commun. 16 (2025) 11303.
[9] F.P. Bundy, H.T. Hall, H.M. Strong, R.H. Wentorf, Nature 176 (1955) 51–55.
[10] S. Ferro, J. Mater. Chem. 12 (2002) 2843–2855.
[11] Y. Gong, D. Luo, M. Choe, Y. Kim, B. Ram, M. Zafari, W.K. Seong, P. Bakharev, M. Wang, I.K. Park, S. Lee, T.J. Shin, Z. Lee, G. Lee, R.S. Ruoff, Nature 629 (2024) 348–354.
[12] J. Zhang, J. Wang, G. Zhang, Z. Huo, Z. Huang, L. Wu, Mater. Des. 237 (2024) 112577.
[13] H.T. Hall, Science 169 (1970) 868–869.
[14] H. Tang, M. Wang, D. He, Q. Zou, Y. Ke, Y. Zhao, Carbon 108 (2016) 1–6.
[15] Q. Tao, X. Wei, M. Lian, H. Wang, X. Wang, S. Dong, T. Cui, P. Zhu, Carbon 120 (2017) 405–410.
[16] J. Guignard, M. Prakasam, A. Largeteau, Materials 15 (2022) 2198.
[17] R.H. Wentorf, R.C. DeVries, F.P. Bundy, Science 208 (1980) 873–880.
[18] H. Katzman, W.F. Libby, Science 172 (1971) 1132–1134.
[19] F.P. Bundy, Nature 241 (1973) 116–118.
[20] M. Seal, Nature 182 (1958) 1264–1267.
[21] M. Seal, Nature 185 (1960) 522–523.
[22] S.N. Monteiro, A.L.D. Skury, M.G. de Azevedo, G.S. Bobrovnitchii, J. Mater. Res. Technol. 2 (2013) 68–74.
[23] Y. Tian, B. Xu, D. Yu, Y. Ma, Y. Wang, Y. Jiang, W. Hu, C. Tang, Y. Gao, K. Luo, Z. Zhao, L.-M. Wang, B. Wen, J. He, Z. Liu, Nature 493 (2013) 385–388.
[24] Y.N. Palyanov, Y.M. Borzdov, I.N. Kupriyanov, A.F. Khohkhryakov, D.V. Nechaev, Sci. Rep. 11 (2021) 8421.
[25] Y. Tian, J. Wang, J. Zhang, S. Guan, L. Zhang, B. Wu, Y. Su, M. Huang, L. Zhou, D. He, Diam. Relat. Mater. 110 (2020) 108158.
[26] A. Biswas, G.A. Alvarez, M. Tripathi, J. Lee, T.S. Pieshkov, C. Li, B. Gao, A.B. Puthirath, X. Zhang, T. Gray, J. Elkins, R. Vajtai, P. Dai, A.G. Birdwell, M.R. Neupane, T. Ivanov, E.J. Garratt, B.B. Pate, A.K. Roy, A. Dalton, Z. Tian, P.M. Ajayan, J. Mater. Chem. C 12 (2024) 3053–3062.
[27] K. Chen, B. Song, N. K. Ravichandran, Q. Zheng, X. Chen, H. Lee, H. Sun, S. Li, G. A. G. U. Gamage, F. Tian, Z. Ding, Q. Song, A. Raj, H. Wu, P. Koirala, A. J. Schmidt, K.

Watanabe, B. Lv, Z. Ren, L. Shi, D. G. Cahill, T. Taniguchi, D. Broido, G. Chen, Science 367 (2020) 555-559.
[28] A.R. Badzian, A. Klokocki, J. Cryst. Growth 52 (1981) 843–847.
[29] A. De Vita, G. Galli, A. Canning, R. Car, Nature 379 (1996) 523–526.
[30] K. Luo, B. Liu, W. Hu, X. Dong, Y. Wang, Q. Huang, Y. Gao, L. Sun, Z. Zhao, Y. Wu, Y. Zhang, M. Ma, X.-F. Zhou, J. He, D. Yu, Z. Liu, B. Xu, Y. Tian, Nature 607 (2022) 486–491.
[31] H. Zhang, Z. Yan, H. Zhang, G. Chen, Funct. Diam. 5 (2025) 2533896.
[32] G. Davies, T. Evans, Proc. A 328 (1972), 413–427.
[33] Y.G. Gogotsi, A. Kailer, K.G. Nickel, Nature 401 (1999) 663–664.
[34] E. O'Bannon, G. Xia, F. Shi, R. Wirth, A. King, L. Dobrzhinetskaya, Diam. Relat. Mater. 108 (2020) 107876.
[35] R. A. Khmelnitsky, A. A. Gippius, Phase Transitions 87 (2013) 175-192.
[36] B.B. Bokhonov, D.V. Dudina, M.R. Sharafutdinov, Diam. Relat. Mater. 118 (2021) 108563.
[37] X. Yan, J. Wei, K. An, J. Liu, L. Chen, Y. Zheng, X. Zhang, C. Li, Diam. Relat. Mater. 120 (2021) 108647.
[38] B. Ali, H. Xu, D. Chetty, R.T. Sang, I.V. Litvinyuk, M. Rybachuk, J. Phys. Chem. Lett. 13 (2022) 2679–2685.
[39] D. Saada, J. Adler, R. Kalish, Int. J. Mod. Phys. C 09 (1998) 61–69.
[40] D.P. Hickey, K.S. Jones, R.G. Elliman, Diam. Relat. Mater. 18 (2009) 1353–1359.
[41] D.J. Erskine, W.J. Nellis, Nature 349 (1991) 317–319.
[42] G.-W. Chen, S.-C. Zhu, L. Xu, Y.-M. Li, Z.-P. Liu, Y. Hou, H. Mao, JACS Au 4 (2024) 3413–3420.
[43] T. Evans, P.F. James, Proc. R. Soc. Lond. Math. Phys. Sci. 277 (1964) 260–269.
[44] S. Eswarappa Prameela, T.M. Pollock, D. Raabe, M.A. Meyers, A. Aitkaliyeva, K.-L. Chintersingh, Z.C. Cordero, L. Graham-Brady, Nat. Rev. Mater. 8 (2023) 81–88.
[45] Z. Li, Y. Wang, M. Ma, H. Ma, W. Hu, X. Zhang, Z. Zhuge, S. Zhang, K. Luo, Y. Gao, L. Sun, A.V. Soldatov, Y. Wu, B. Liu, B. Li, P. Ying, Y. Zhang, B. Xu, J. He, D. Yu, Z. Liu, Z. Zhao, Y. Yue, Y. Tian, X. Li, Nat. Mater. 22 (2023) 42–49.
[46] P. Németh, K. McColl, L.A.J. Garvie, C.G. Salzmann, M. Murri, P.F. McMillan, Nat. Mater. 19 (2020) 1126–1131.
[47] Y. Yue, Y. Gao, W. Hu, B. Xu, J. Wang, X. Zhang, Q. Zhang, Y. Wang, B. Ge, Z. Yang, Z. Li, P. Ying, X. Liu, D. Yu, B. Wei, Z. Wang, X.-F. Zhou, L. Guo, Y. Tian, Nature 582 (2020) 370–374.
[48] B. Li, P. Ying, Y. Gao, W. Hu, L. Wang, Y. Zhang, Z. Zhao, D. Yu, J. He, J. Chen, B. Xu, Y. Tian, Nano Lett. 22 (2022) 4979–4984.
[49] A.C. Li, B. Li, F. González-Cataldo, R.E. Rudd, B. Militzer, E.M. Bringa, M.A. Meyers, Mater. Sci. Eng. R Rep. 161 (2024) 100857.
[50] A.B. Peters, D. Zhang, S. Chen, C. Ott, C. Oses, S. Curtarolo, I. McCue, T.M. Pollock, S. Eswarappa Prameela, Nat. Commun. 15 (2024) 3328.
[51] J. Qian, C. Pantea, J. Huang, T.W. Zerda, Y. Zhao, Carbon 42 (2004) 2691–2697.
[52] H. Wei, H. Zhan, D. Legut, S. Zhang, Carbon 235 (2025) 120079.
[53] R. Qi, R. Shi, Y. Li, Y. Sun, M. Wu, N. Li, J. Du, K. Liu, C. Chen, J. Chen, F. Wang, D. Yu, E.-G. Wang, P. Gao, Nature 599 (2021) 399–403.

[54] H.G. Johnson, S.P. Bennett, R. Barua, L.H. Lewis, D. Heiman, Phys. Rev. B 82 (2010) 085202.
[55] C.E. Anderson, T.G. Trucano, S.A. Mullin, Int. J. Impact Eng. 9 (1990) 89–113.
[56] A.J. Piekutowski, Int. J. Impact Eng. 14 (1993) 573–586.
[57] B.G. Cour-Palais, Int. J. Impact Eng. 5 (1987) 221–237.
[58] Y. Zhou, H. Tian, F. Li, P. Zuo, Y. Ma, Nanoscale 18 (2026) 8475–8499.
[59] X. Yan, M.A. Firestone, M. Keçeli, S. Chaudhuri, E. Huerta, Carbon 252 (2026) 121366.
[60] R. Narulkar, S. Bukkapatnam, L.M. Raff, R. Komanduri, Comput. Mater. Sci. 45 (2009) 358–366.
[61] S. Fahy, S.G. Louie, M.L. Cohen, Phys. Rev. B 34 (1986) 1191–1199.
[62] X. Dong, X.-F. Zhou, G.-R. Qian, Z. Zhao, Y. Tian, H.-T. Wang, J. Phys. Condens. Matter 25 (2013) 145402.
[63] R. Z. Khaliullin, H. Eshet, T. D. Kühne, J. Behler, M. Parrinello, Nat. Mater. 10, 693-697 (2011).
[64] L.C. Liang, X-Ray Spectrom. 21 (1992) 191–192.
[65] M.N.R. Ashfold, J.P. Goss, B.L. Green, P.W. May, M.E. Newton, C.V. Peaker, Chem. Rev. 120 (2020) 5745–5794.
[66] J.L. Pouchou, F. Pichoir, Scanning Microsc. 1993 (1993) 167-189.
[67] J.A. Rogers, N. Bass, P.T. Mead, A. Mote, G.D. Lukasik, M. Intardonato, K. Harrison, J.D. Leaverton, K.R. Kota, J.W. Wilkerson, J.N. Reddy, W.D. Kulatilaka, T.E. Lacy Jr., Rev. Sci. Instrum. 93 (2022) 085106.
[68] G. Henkelman, B.P. Uberuaga, H. Jónsson, J. Chem. Phys. 113 (2000) 9901–9904.
[69] B. Hourahine, B. Aradi, V. Blum, F. Bonafé, A. Buccheri, C. Camacho, C. Cevallos, M.Y. Deshaye, T. Dumitrică, A. Dominguez, S. Ehlert, M. Elstner, T. van der Heide, J. Hermann, S. Irle, J.J. Kranz, C. Köhler, T. Kowalczyk, T. Kubař, I.S. Lee, V. Lutsker, R.J. Maurer, S.K. Min, I. Mitchell, C. Negre, T.A. Niehaus, A.M.N. Niklasson, A.J. Page, A. Pecchia, G. Penazzi, M.P. Persson, J. Řezáč, C.G. Sánchez, M. Sternberg, M. Stöhr, F. Stuckenberg, A. Tkatchenko, V.W. -z. Yu, T. Frauenheim, J. Chem. Phys. 152 (2020) 124101.
[70] A. Hjorth Larsen, J. Jørgen Mortensen, J. Blomqvist, I.E. Castelli, R. Christensen, M. Dułak, J. Friis, M.N. Groves, B. Hammer, C. Hargus, E.D. Hermes, P.C. Jennings, P. Bjerre Jensen, J. Kermode, J.R. Kitchin, E. Leonhard Kolsbjerg, J. Kubal, K. Kaasbjerg, S. Lysgaard, J. Bergmann Maronsson, T. Maxson, T. Olsen, L. Pastewka, A. Peterson, C. Rostgaard, J. Schiøtz, O. Schütt, M. Strange, K.S. Thygesen, T. Vegge, L. Vilhelmsen, M. Walter, Z. Zeng, K.W. Jacobsen, J. Phys. Condens. Matter 29 (2017) 273002.
[71] S. Smidstrup, A. Pedersen, K. Stokbro, H. Jónsson, J. Chem. Phys. 140 (2014) 214106.
[72] S. Plimpton, J. Comput. Phys. 117 (1995) 1–19.
[73] A.C.T. van Duin, S. Dasgupta, F. Lorant, W.A. Goddard, J. Phys. Chem. A 105 (2001) 9396–9409.
[74] L. Verlet, Phys. Rev. 159 (1967) 98–103.
[75] S. Nosé, Mol. Phys. 52 (1984) 255–268.
[76] W.G. Hoover, Phys. Rev. A 31 (1985) 1695–1697.
[77] J. D. Weeks, D. Chandler, H. C. Anderson, J. Chem. Phys. 54, (1971), 5237–5247.
[78] A. Ahmed, R.J. Sadus, Phys. Rev. E 80 (2009) 061101.

Supplementary material for

# Diamond-to-graphite transformation under hypersonic impact

**The PDF file includes:**

Figures **S1** to **S31**
Tables **S1** to **S2**
Supplementary Text

**Other Supplementary material for this manuscript includes:**

Movies **S1** to **S7**

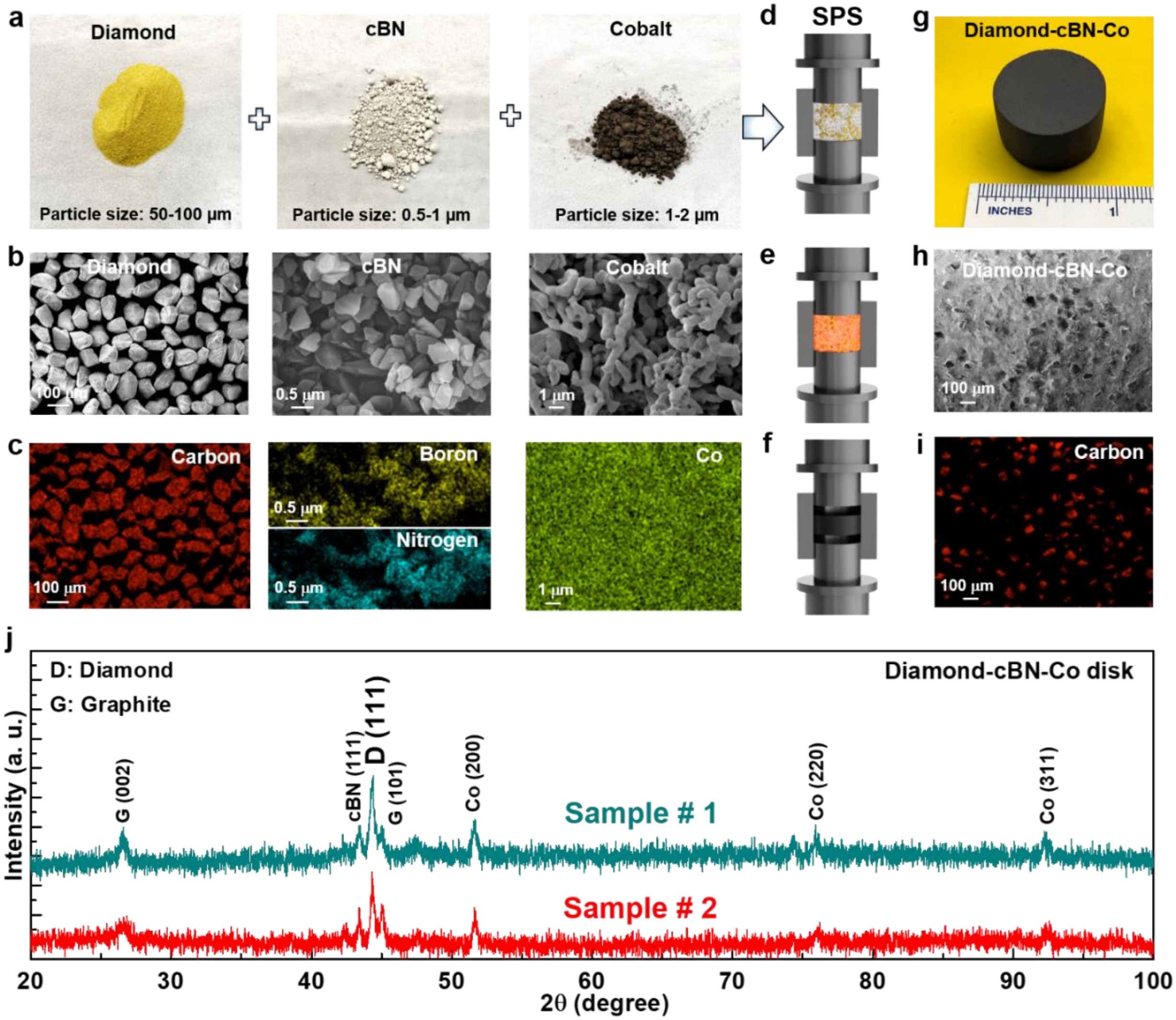

**Fig. S1. Synthesis of diamond-cBN-Co composite. (a)** Commercially available diamond, cubic boron nitride (cBN), and cobalt (Co) powders with different particle sizes are mixed homogeneously in a mortar pestle for 15 min before placing them into the SPS chamber. Diamond powder is yellow due to the presence of nitrogen, causing the diamond to reflect yellow light [10,51]. **(b)**, **(c)** Corresponding FESEM images and EDS elemental mapping show different grain sized particles. **(d), (e), (f)** Schematic of the SPS process which was performed at 1400 °C, 90 MPa pressure for 1 hr in high purity 5N Argon atmosphere. **(g), (h), (i)** SPS produced black colored disk, and its FESEM surface morphology and EDS elemental mapping of carbon showing the presence of diamond particles throughout the sample. **(j)** Wider 2θ range XRD scans of two different samples sintered at the same conditions.

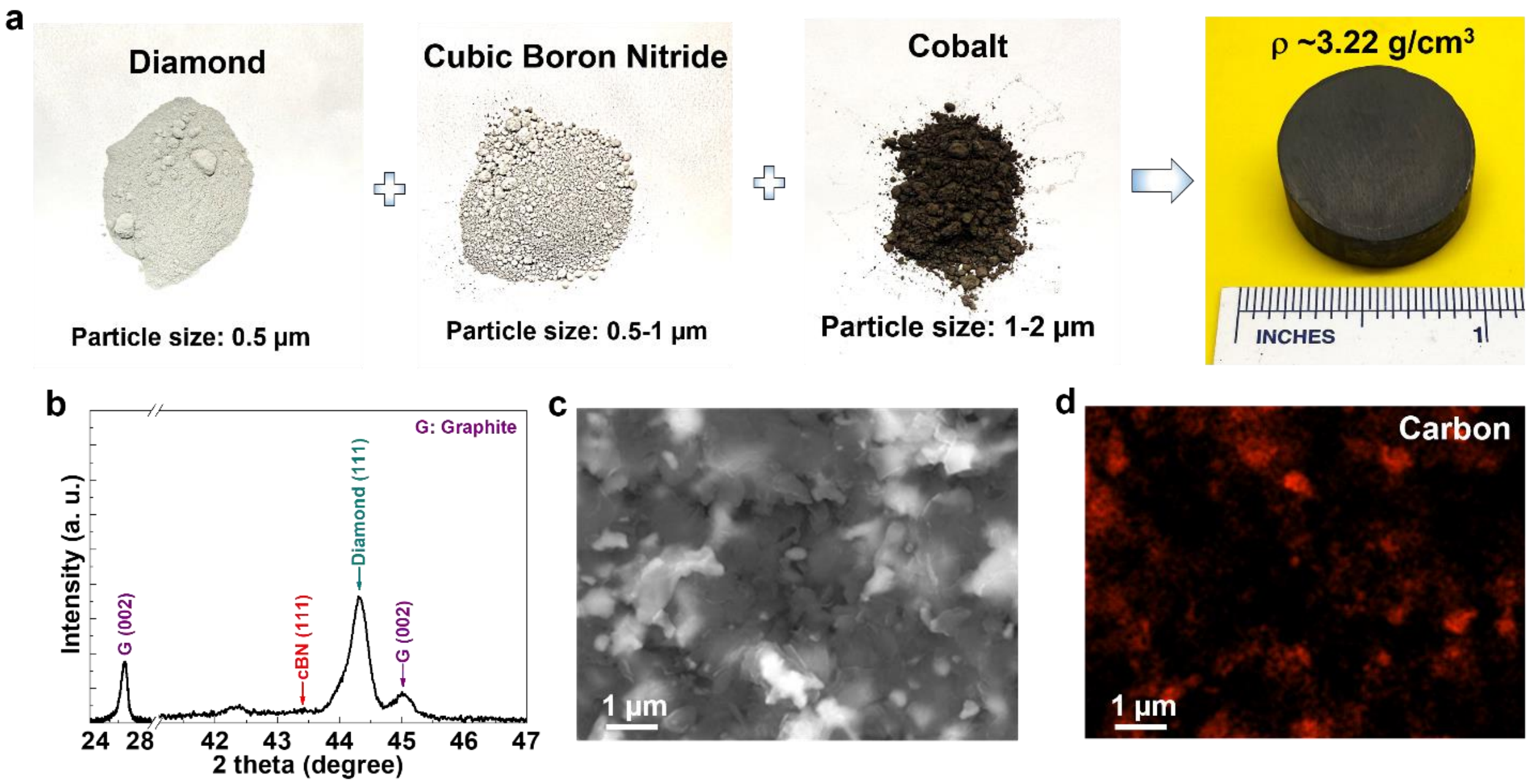

**Fig. S2. Synthesis attempt of diamond-cBN-Co composites by using smaller grain sizes of diamond particles. (a)** Respective diamond, cBN, and cobalt powders and a sintered disk. **(b)** Corresponding XRD pattern showing the strong graphite (002) peak. **(c), (d)** FESEM surface morphology and EDS elemental map of carbon is shown. The composite was sintered at the same SPS conditions of 1400 °C, 90 MPa pressure for 1 hr.

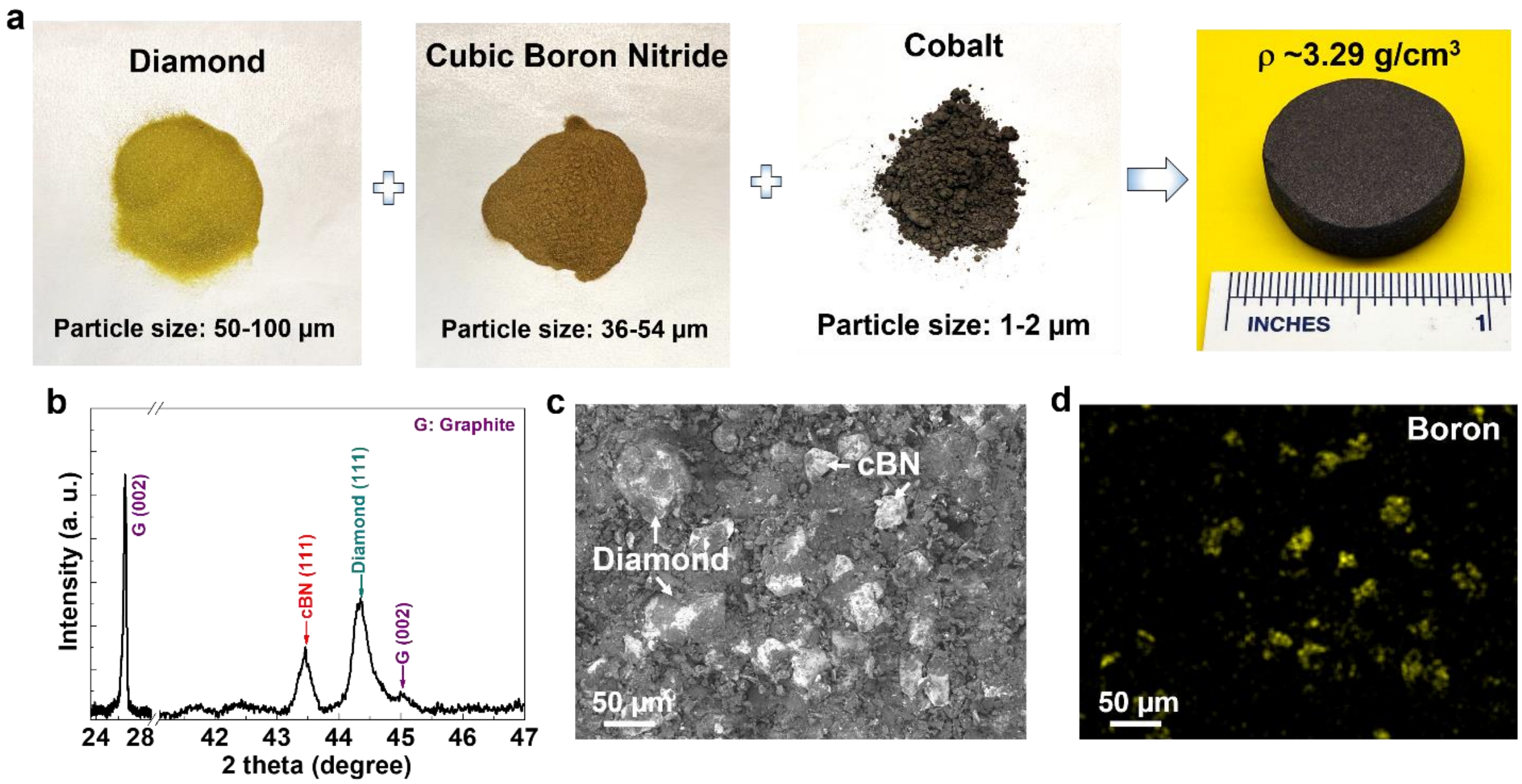

**Fig. S3. Synthesis attempt of diamond-cBN-Co composites by using larger grain sizes of cBN particles.** **(a)** Respective diamond, cBN, and cobalt powders and sintered disk. The larger grain sizes cBN powder is yellowish due to defects. **(b)** Corresponding XRD pattern showing the intense graphite (002) peak. **(c)** FESEM surface morphology shows the presence of the diamond and cBN particles. **(d)** EDS elemental map of boron confirms the presence of cBN particles. The composite was sintered at the same SPS conditions of 1400 °C, 90 MPa pressure for 1 hr.

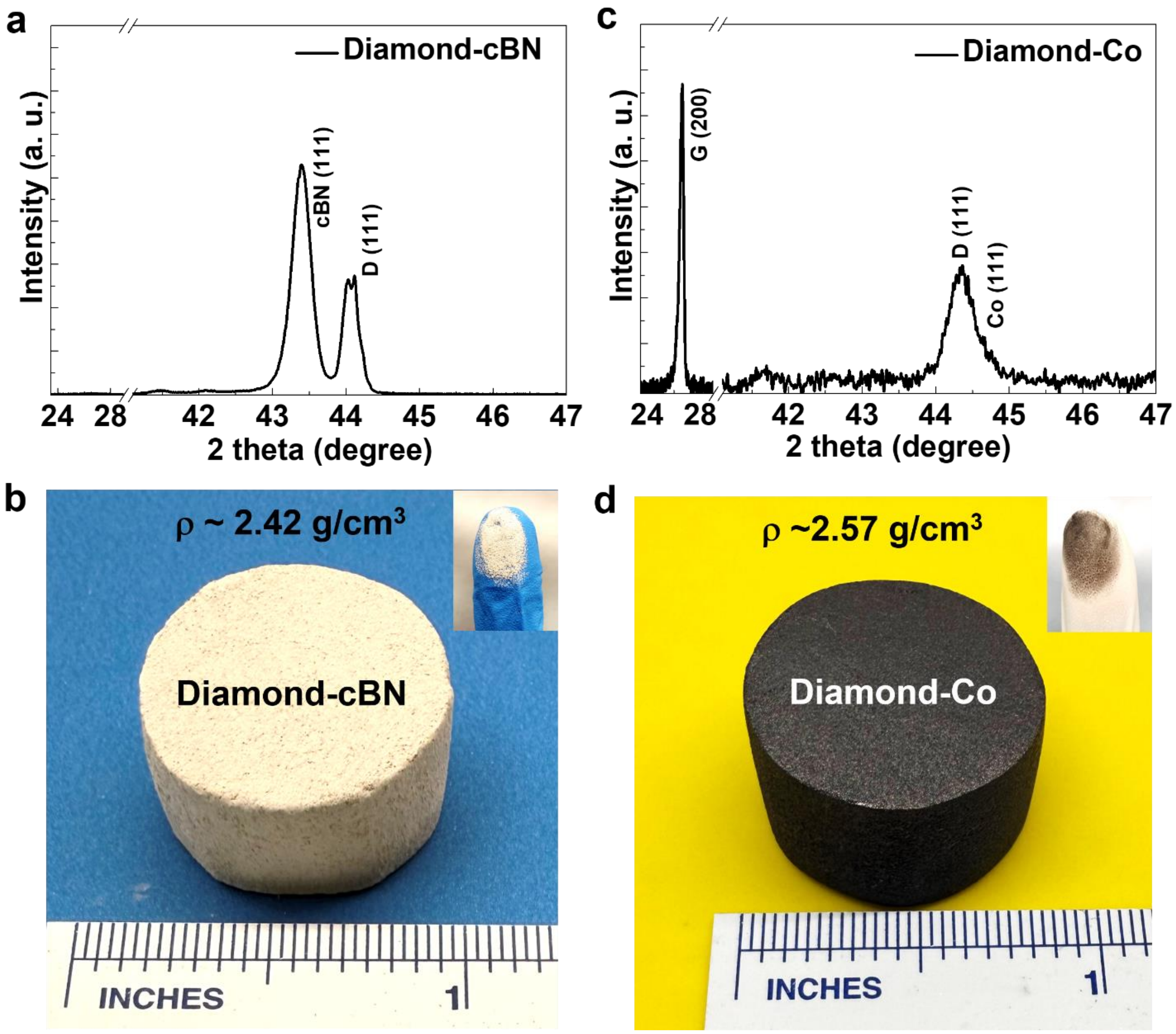

**Fig. S4. Synthesis attempts of the diamond-based composite using diamond-cBN or diamond-Co. (a), (b)** Synthesis attempt of diamond-cBN composite (without Co) producing an ultra-soft disk (almost non-sintered) which can be easily broken (rubbed) by pressing with finger tips (upper right inset of **b**). **(c), (d)** Synthesis of diamond-Co composite (without cBN) which is comparatively harder, however shows the presence of a more intense graphitic peak than diamond. Again, the sintering conditions are the same of 1400 °C, 90 MPa pressure, for 1hr, as used for the main diamond-cBN-Co composite.

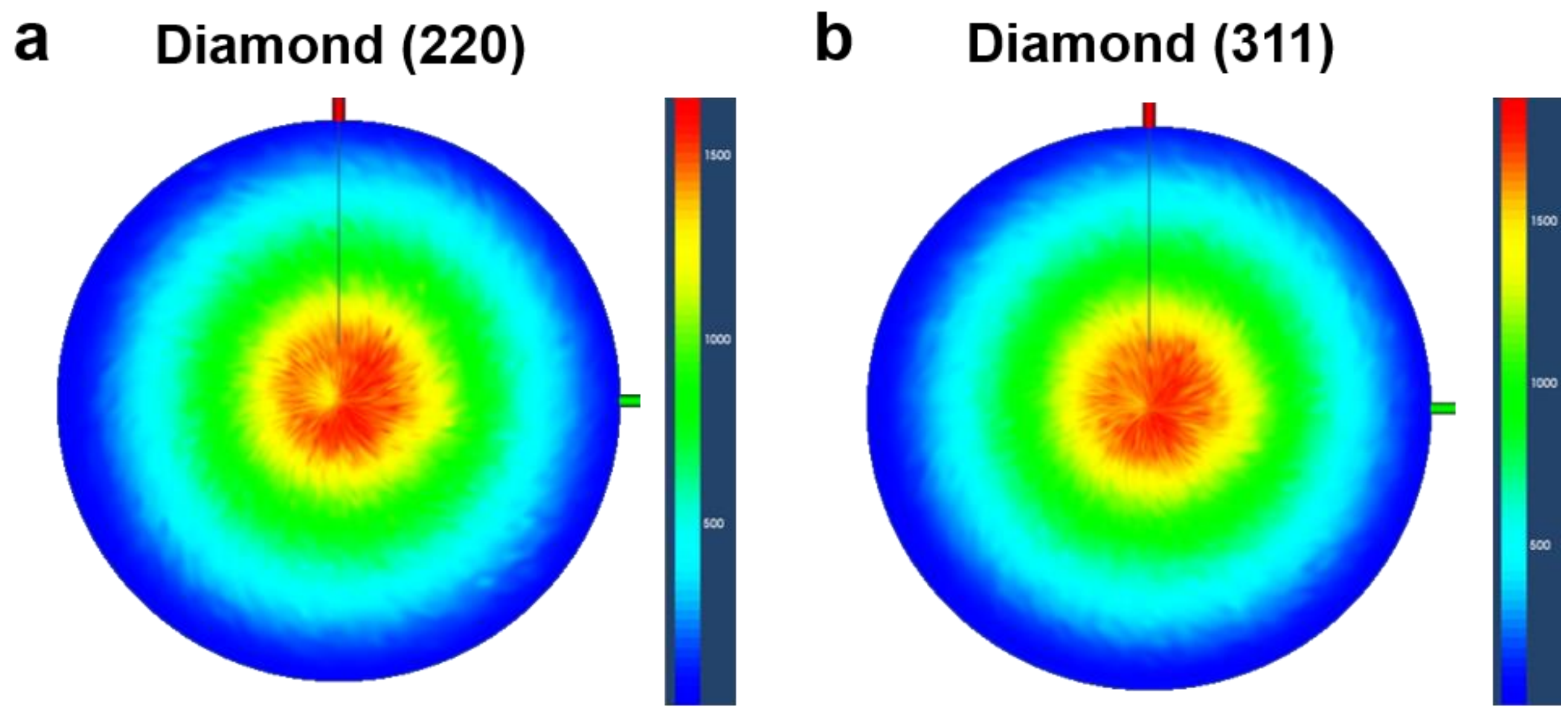

**Fig. S5. Pole figure mapping of the composites. (a), (b)** XRD pole figure map of the (220) and (311) Bragg peaks of diamond show no grain-like features distributed throughout the sample (as like the (111) peak pole figure case, main **Fig. 1b)**, possibly having very small numbers of facets with relatively lower intensity.

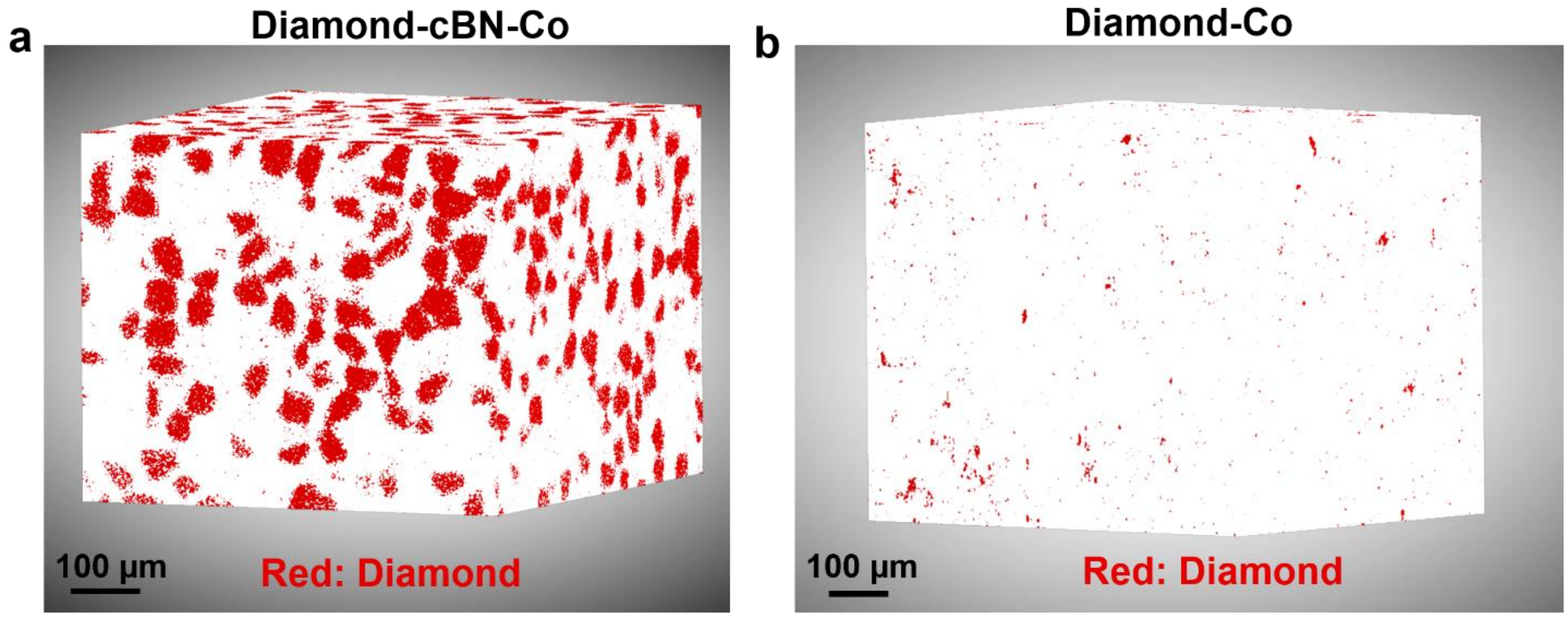

**Fig. S6. X-ray microscopy imaging of the composite. (a)** X-ray microscopy imaging of diamond particles (red) in the three-phase diamond-cBN-Co composite. **(b)** In comparison, X-ray microscopy imaging of the diamond particles (red) in the diamond-Co composite (without cBN), showing smaller particle sizes with a lower volume of diamond. Here we kept non-diamond regions white for better visualization of diamonds. This shows that addition of cBN as a matrix helps in stabilizing diamond.

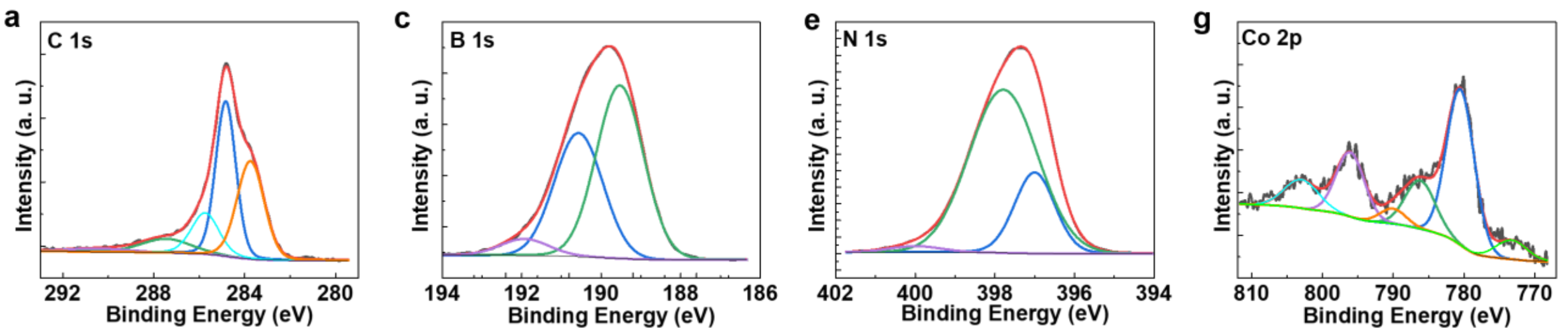

**b**

| Position (eV) | Peak | Area (%) |
|---|---|---|
| 283.74 | Metal-C, B-C | 33.59 |
| 284.83 | $sp^3$ C | 39.88 |
| 285.74 | C=C, C=O | 15.18 |
| 287.46 | C-N | 8.84 |
| 290.48 | O-C=O | 2.52 |

**d**

| Position (eV) | Peak | Area (%) |
|---|---|---|
| 189.53 | B-C | 54.97 |
| 190.57 | B-N | 39.49 |
| 192.93 | B-O | 5.54 |

**f**

| Position (eV) | Peak | Area (%) |
|---|---|---|
| 397.00 | C=N-C | 22.67 |
| 397.79 | B-N | 75.13 |
| 399.99 | C-N | 2.24 |

**h**

| Position (eV) | Peak | Area (%) |
|---|---|---|
| 773.30 | $Co^0$ | 6.16 |
| 780.50 | $Co^{3+}$ | 45.37 |
| 786.18 | Satellite | 15.16 |
| 790.03 | Satellite | 4.06 |
| 796.06 | $Co^{3+}$ | 19.09 |
| 802.98 | Satellite | 10.17 |

**Fig. S7. XPS of the sintered pristine three-phase diamond-cBN-Co disk.** Core-level XPS scans and its fitting show various possible bonding for **(a), (b)** C 1s, **(c), (d)** B 1s, **(e), (f)** N 1s, and **(g), (h)** Co 2p**.**

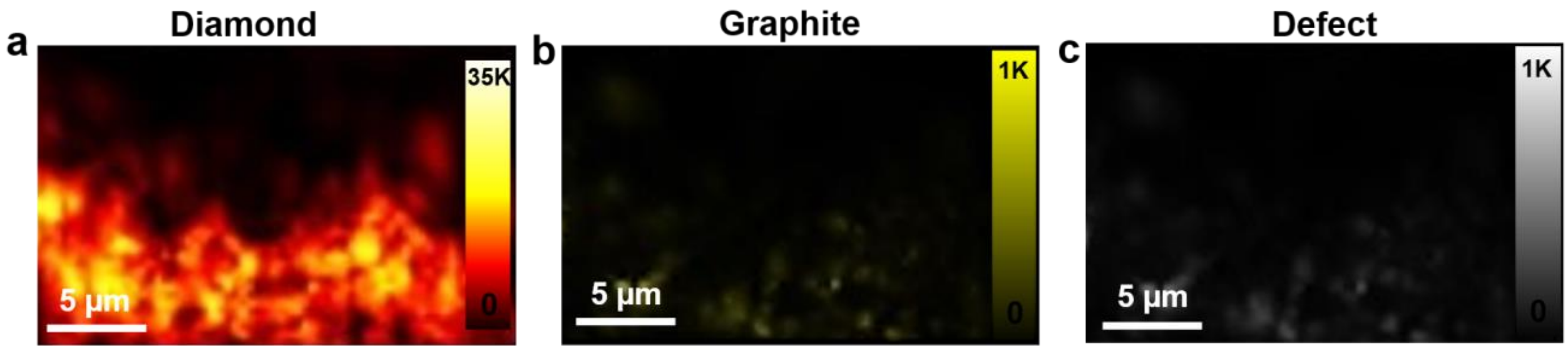

**Fig. S8. Raman intensity map of the sintered pristine three-phase diamond-cBN-Co disk. (a), (b), (c)** Raman intensity map from some portion of a diamond particle showing strong intensity from diamond, and almost no signal from graphite or defects for pristine sample.

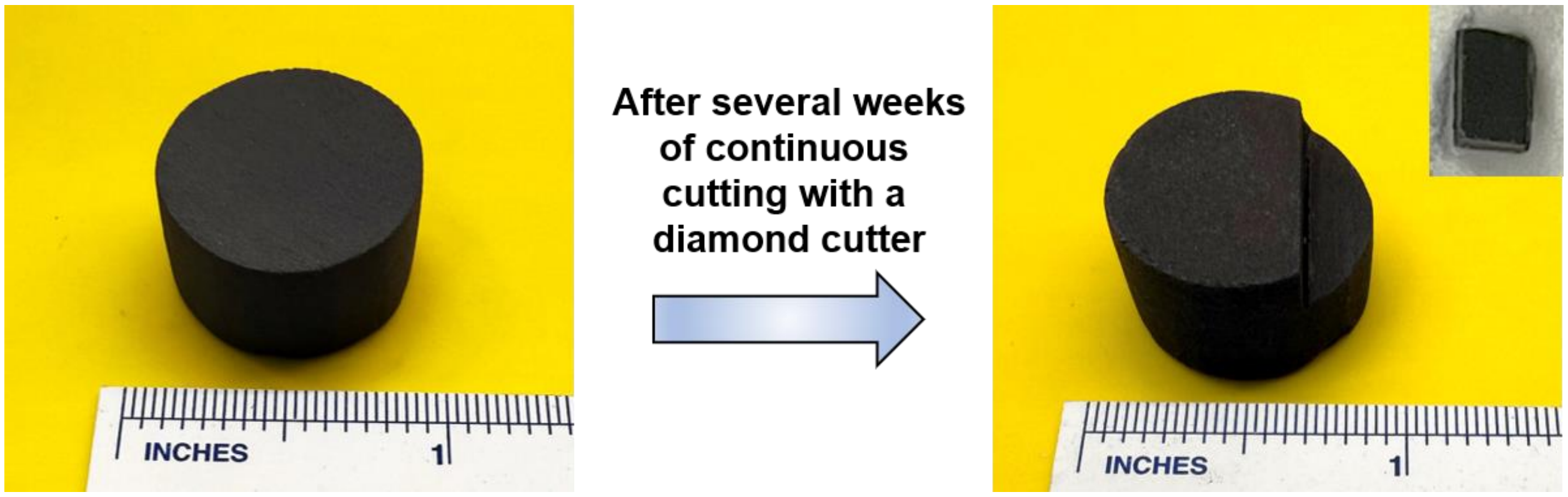

**Fig. S9. Machining of the diamond-cBN-Co composite.** A low-speed diamond saw (Model 650, South Bay Technology Inc.) with dimensions of 13″ W × 9″ H × 13″ D and a maximum wheel speed of 300 rpm was used to cut the composite disk. A rectangular sample piece of about 3×2 $mm^2$ (inset) was obtained after continuous cutting for eight days. During cutting, the diamond wheel operated at its maximum speed (300 rpm) while fully submerged and rotating through a bath of isopropanol to minimize friction. The pieces were subsequently used for mechanical, electrical, and magnetization measurements.

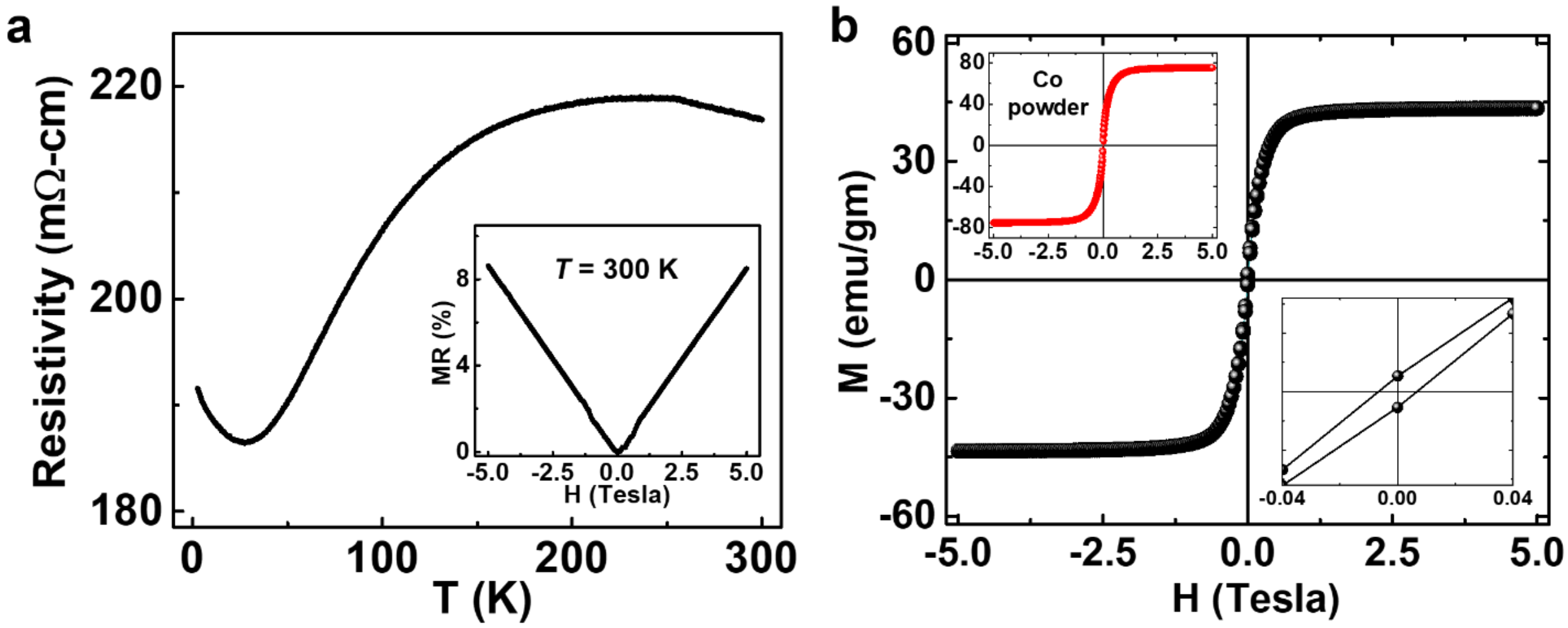

**Fig. S10. Electrical transport and magnetic properties of the diamond-cBN-Co composite. (a)** Composite is electrically conducting with room temperature resistivity of ~217 mΩ-cm. Inset shows the V-shape linear field-dependent magnetoresistance. **(b)** It shows ferromagnetic behavior at room temperature with a hysteretic M-H loop with a low coercivity of ~656 Oe, shown in lower-right inset. Upper-left inset shows the ferromagnetic loop of pure cobalt (Co) powder.

| Projectile material type | Specific projectile material | Projectile diameter (mm) | Sabot (mm) | L/D ratio | Projectile mass (g) | Projectile package mass (g) | Projectile shape | Projectile speed (m/s) | Projectile KE (kJ) | Target temperature (°C) | Room temperature (°C) | Room humidity (%) |
|---|---|---|---|---|---|---|---|---|---|---|---|---|
| Aluminum (Al) | 2017-T4 Aluminum | BS-1.00 | 6 | 1 | 0.049 | 2.075 | Sphere (S) | 2584.7 (Mach 7.5) | 0.16 | 21.2 | 21.1 | 57 |
| Aluminum (Al) | 2017-T4 Aluminum | 4 | 4 | 1 | 0.091 | 2.136 | Sphere (S) | 2900.8 (Mach 8.45) | 0.38 | 21.2 | 21.1 | 57 |

**Fig. S11. Parameters used for the impact test.** Impact tests with several ∅1 mm (upper panel) and with a ∅4 mm (lower panel) aluminum metal balls as projectiles.

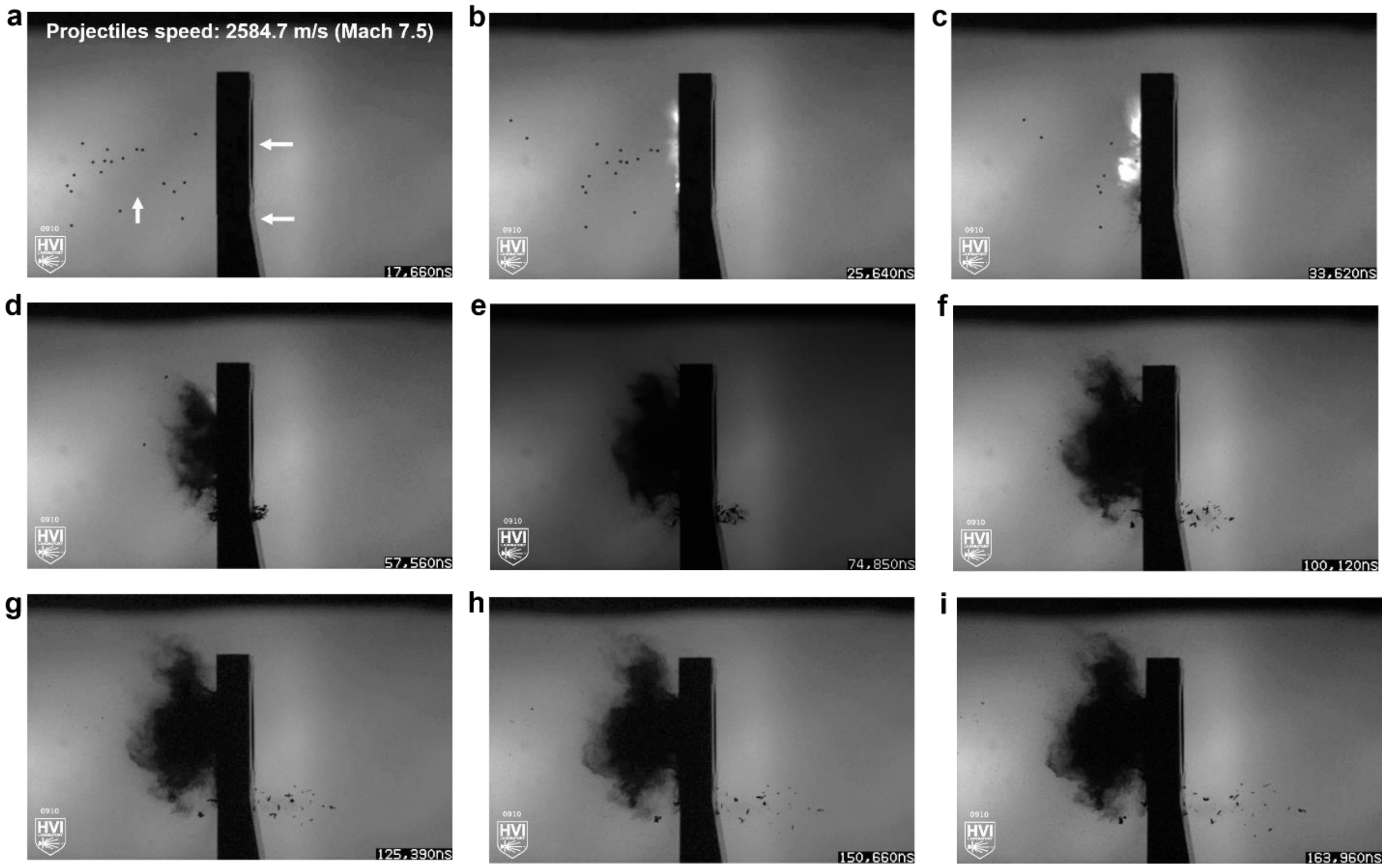

**Fig. S12. Shadowgraphs of the impact test with ∅1 mm projectile (speed 2584.7 m/s, Mach 7.5). (a)-(i)** High-contrast shadowgraphs of the impact event at different time-scales with several ∅1 mm aluminum spherical projectile.

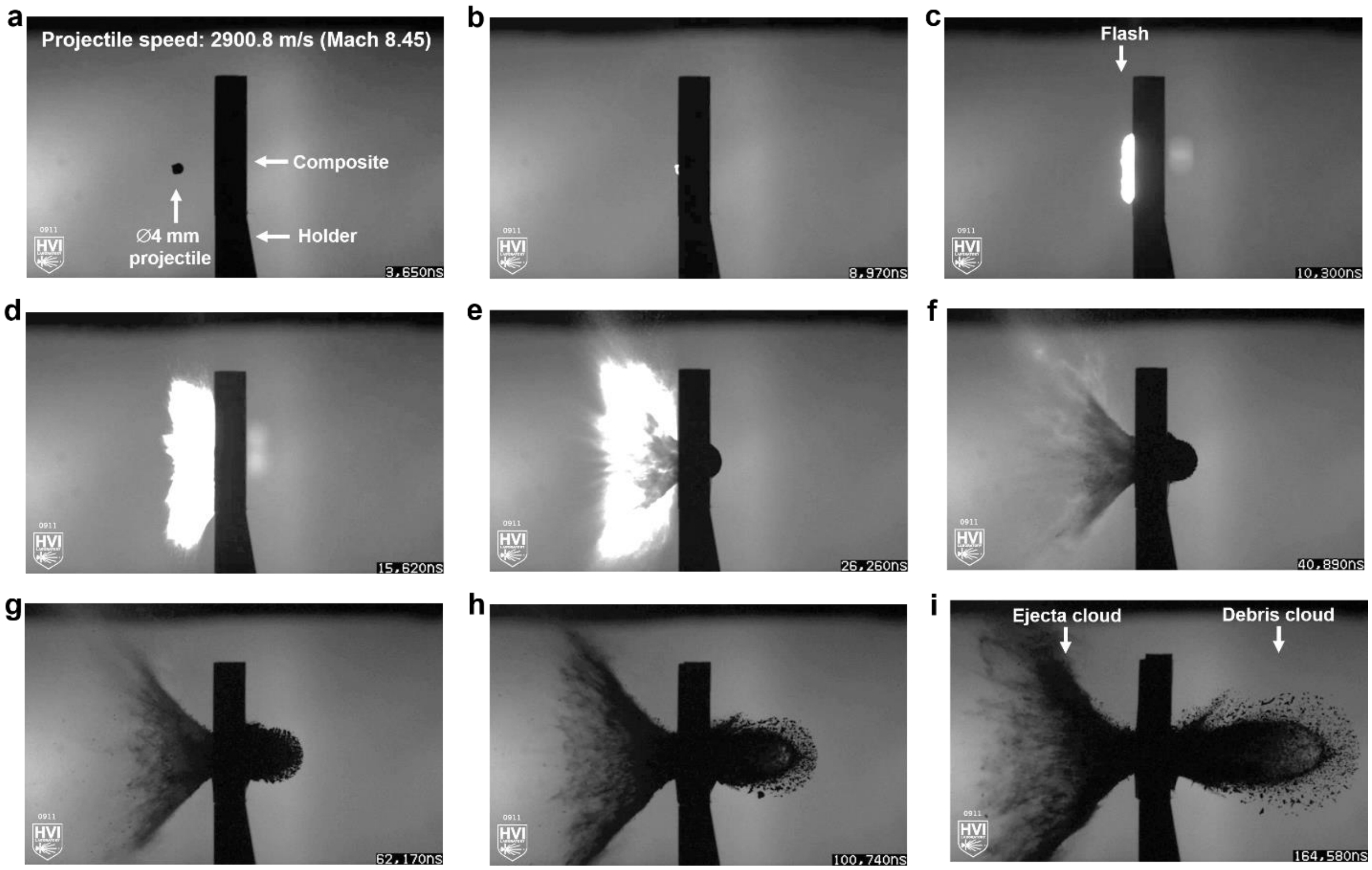

**Fig. S13. Shadowgraphs of the impact test with ∅4 mm projectile (speed 2900.8 m/s, Mach 8.45). (a)-(i)** High-contrast shadowgraphs of the impact event at different time-scales with one ∅4 mm aluminum spherical projectile.

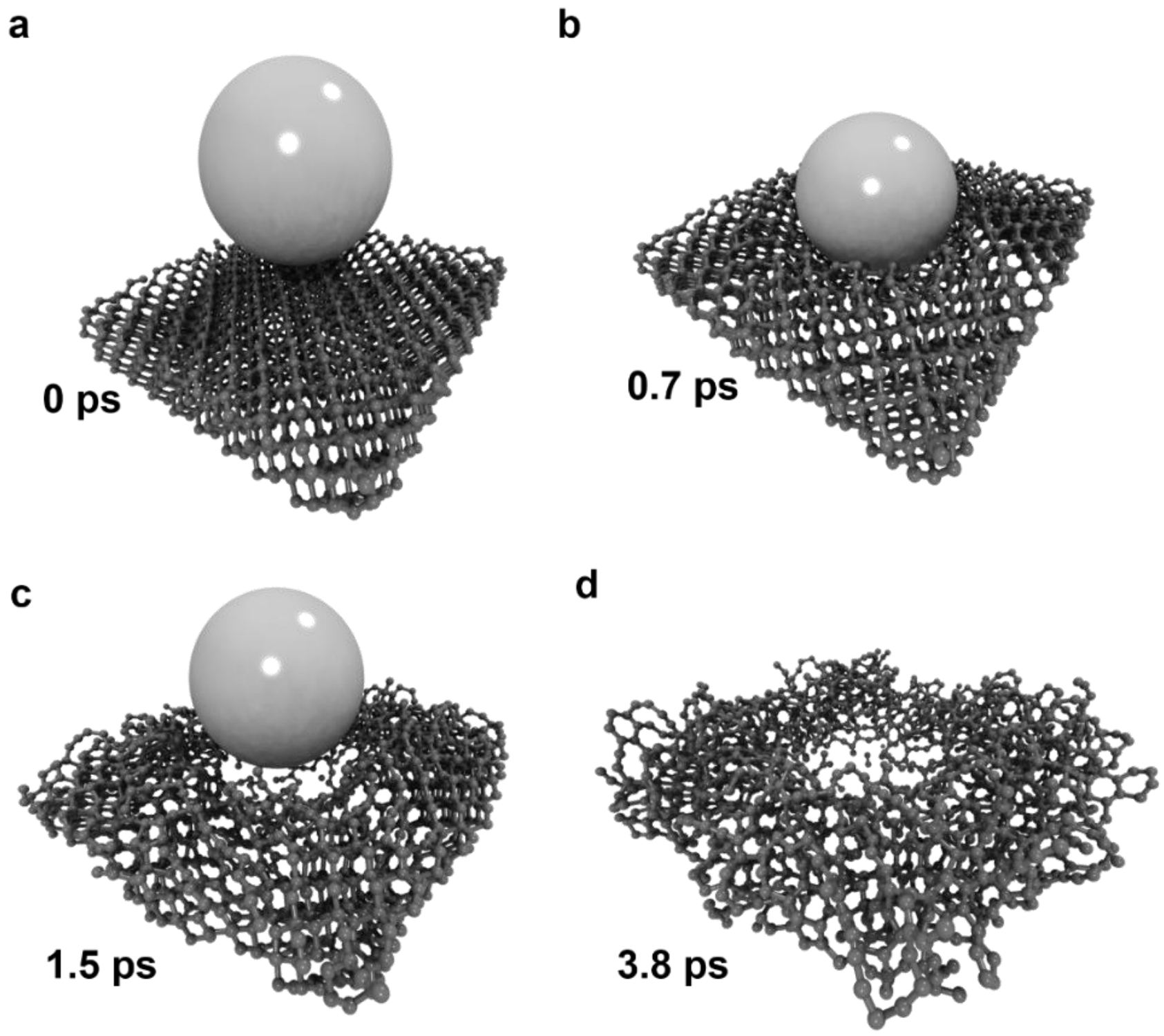

**Fig. S14. Time evolution of the projectile's impact on the diamond surface. (a), (b), (c), (d)** Time evolution of the impacted diamond surface at different simulation times: 0 ps, 0.7 ps, 1.5 ps, and 3.8 ps. At the initial stage (0 ps), the system exhibits a perfectly ordered crystalline structure, corresponding to the equilibrium state after thermal relaxation. At 0.7 ps, the collision and the subsequent penetration phase occur. The projectile's kinetic energy is rapidly converted into potential and thermal energy, generating a compressive shock wave that propagates through the diamond lattice. This initial compression induces the formation of local defects and atomic displacements, particularly in regions of high stress accumulation. At 1.5 ps, the target surface exhibits significant plastic deformation and increased local roughness. The energy transfer leads to the breaking of covalent $sp^3$ bonds and to a partial conversion to $sp^2$ ones, resulting in a structurally disordered region. The accumulation of shear stresses around the impact zone promotes the nucleation of amorphous regions and the loss of crystalline symmetry. By 3.8 ps, the diamond undergoes a pronounced structural transition: the material surrounding the crater exhibits a partial collapse of the lattice, forming an amorphous, highly damaged region, while the inner layers retain minor remnants of tetrahedral order and a considerable fraction of $sp^2$-hybridized carbon atoms.

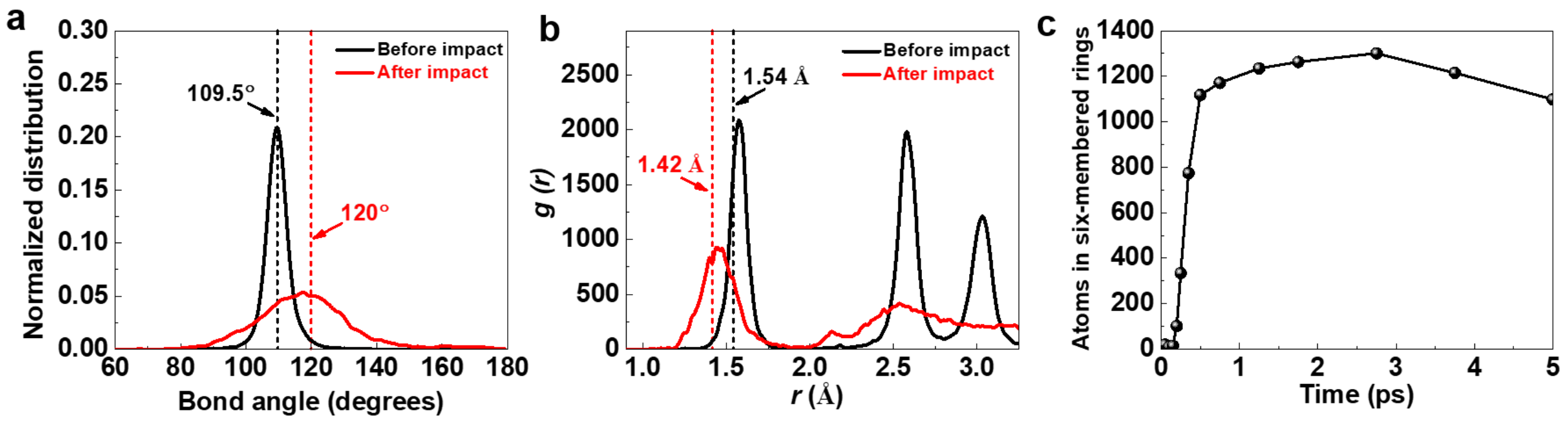

**Fig. S15. Structural signatures of the impact-induced diamond-to-graphite transformation, obtained from the atomic positions. (a)** Bond-angle distribution before and after the impact, with the tetrahedral (109.5°) and trigonal (120°) reference values indicated. **(b)** Radial distribution function before and after the impact, with the C-C distances of diamond (1.54 Å) and graphite (1.42 Å) indicated. **(c)** Number of target carbon atoms belonging to at least one six-membered ring as a function of time, referred to the instant of contact.

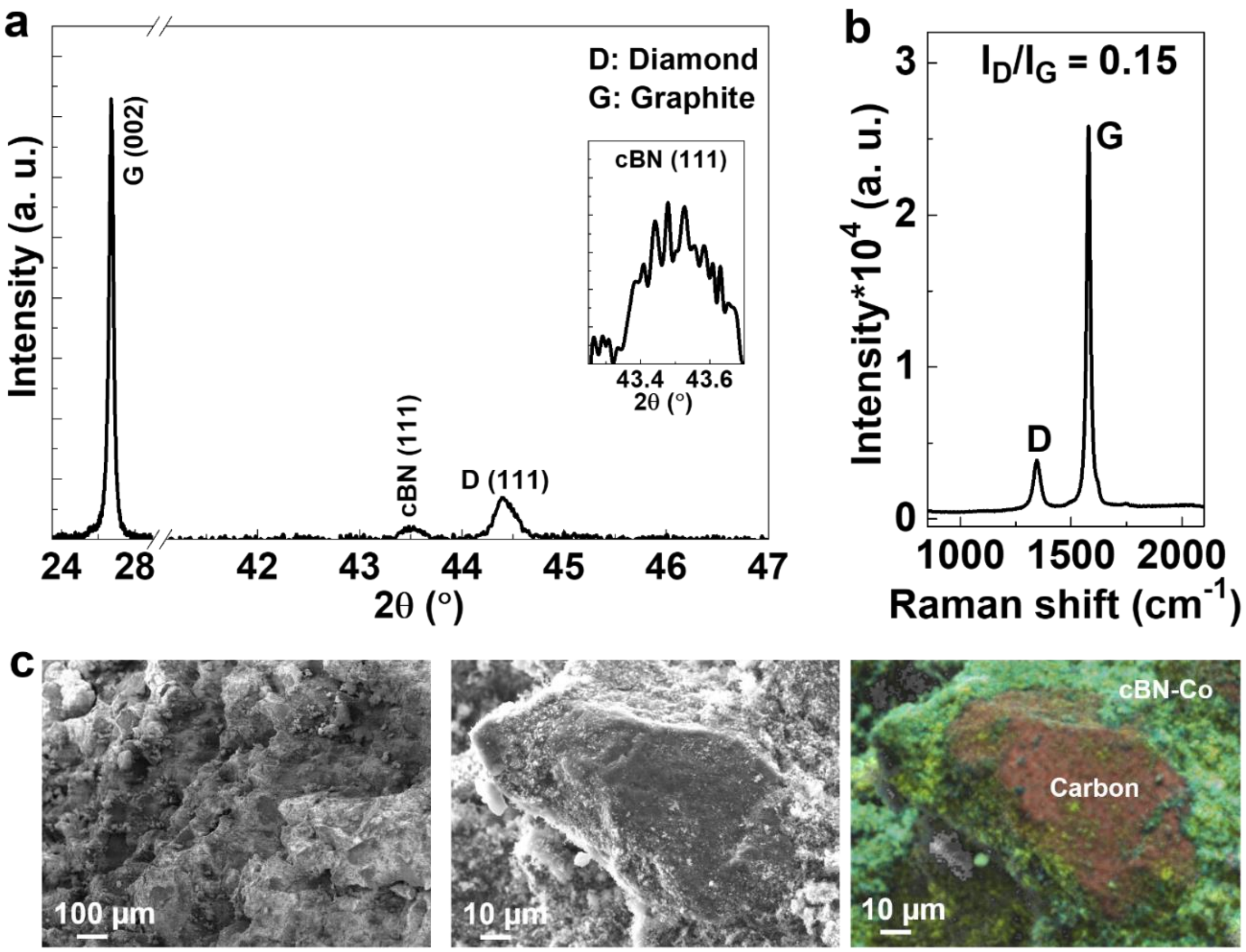

**Fig. S16. Characterizations of the composite after the ∅1 mm projectile impact test. (a)** XRD shows the formation of graphite along with the presence of cBN and diamond phases. **(b)** Raman spectroscopy also shows the graphitization with disorder. **(c)** FESEM morphology and the elemental map showing carbon and cBN-Co regions.

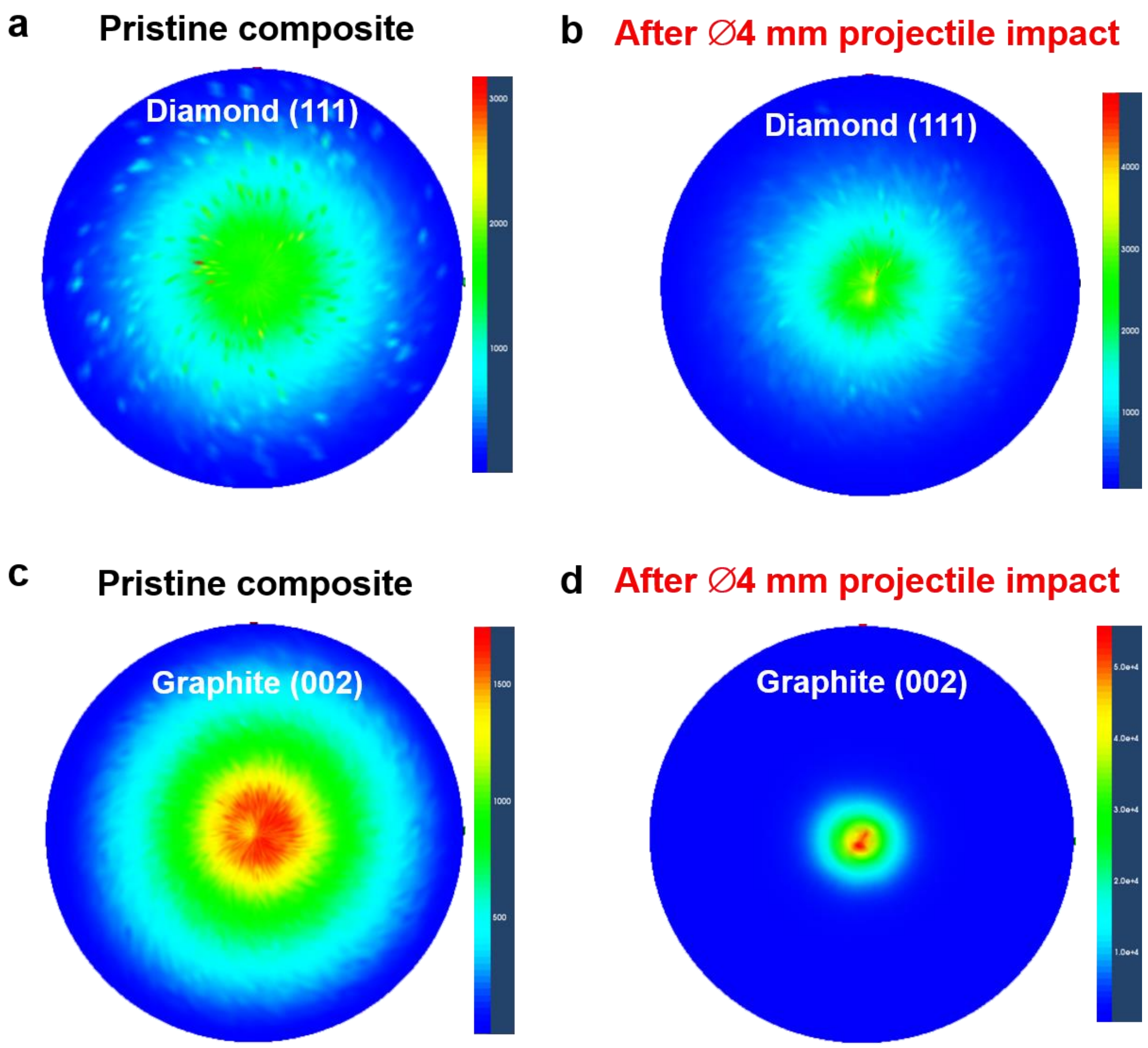

**Fig. S17. Pole figure map of the composite after the ∅4 mm projectile impact test. (a), (b)** XRD pole figure map of the (111) diamond peak. **(c), (d)** XRD pole figure map of the (002) graphite peak of the composite. For pristine composite, the diamond particles are oriented with (111) facets along random directions. In contrast, phase transformed graphite becomes mostly oriented along (002) after the impact.

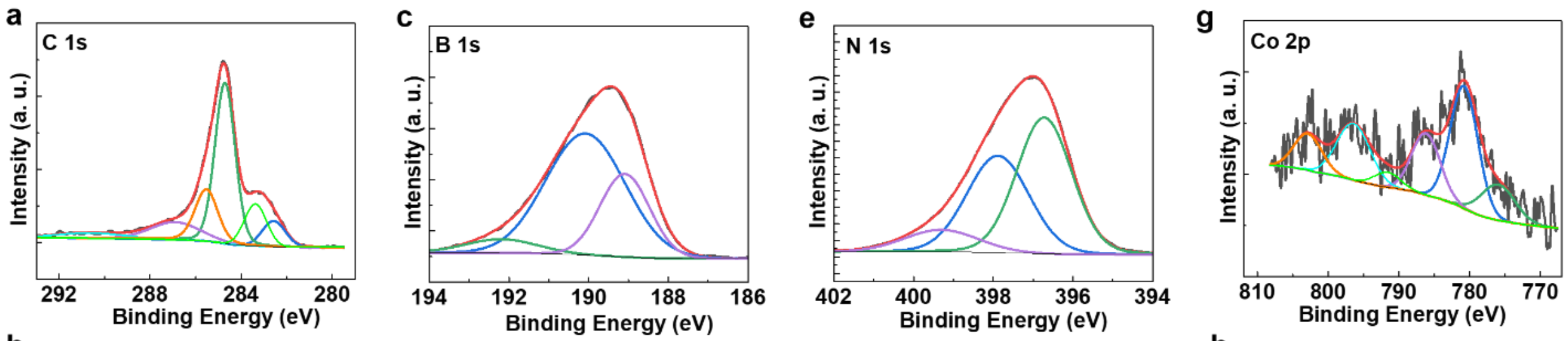

**b**

| Position (eV) | Peak | Area (%) |
|---|---|---|
| 282.57 | Free C | 7.85 |
| 283.38 | Metal-C, B-C | 12.34 |
| 284.71 | sp$^2$ C | 45.59 |
| 285.53 | C=O | 17.36 |
| 286.90 | C-O | 12.23 |
| 290.58 | O-C=O | 4.63 |

**d**

| Position (eV) | Peak | Area (%) |
|---|---|---|
| 189.09 | B-C | 27.92 |
| 190.08 | B-N | 65.50 |
| 192.19 | B-O | 6.58 |

**f**

| Position (eV) | Peak | Area (%) |
|---|---|---|
| 396.72 | C=N-C | 50.03 |
| 397.88 | B-N | 39.63 |
| 399.31 | C-N | 10.34 |

**h**

| Position (eV) | Peak | Area (%) |
|---|---|---|
| 775.95 | $Co^0$ | 12.56 |
| 780.79 | $Co^{3+}$ | 35.30 |
| 786.18 | Satellite | 18.22 |
| 791.59 | Satellite | 4.03 |
| 796.43 | $Co^{3+}$ | 19.35 |
| 802.88 | Satellite | 10.55 |

**Fig. S18. XPS of the composite after the ∅4 mm projectile impact test.** Core-level elemental XPS sans and their fittings show various possible bonding for **(a), (b)** C 1s, **(c), (d)** B 1s, **(e), (f)** N 1s, and **(g), (h**) Co 2p**.**

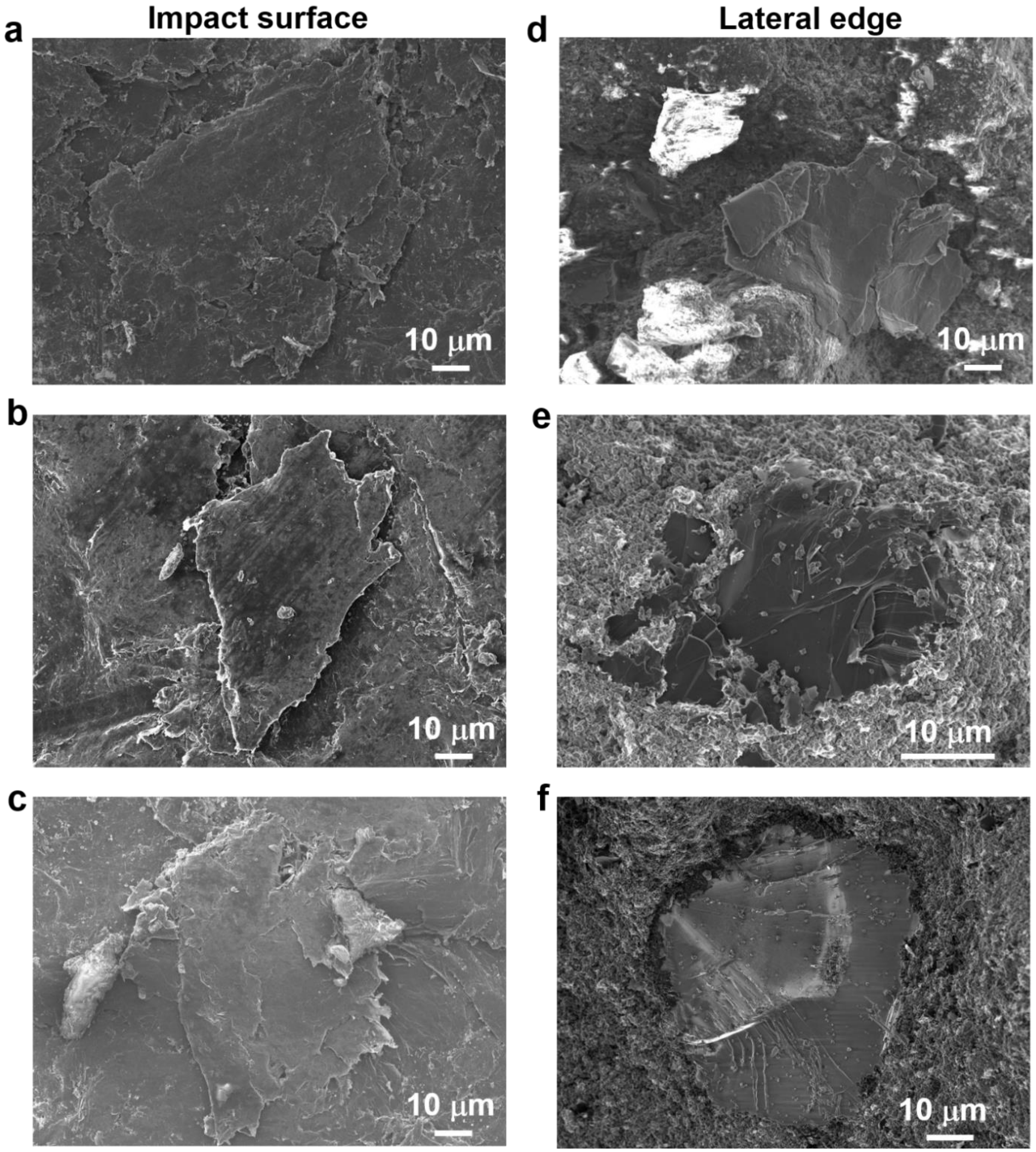

**Fig. S19. FESEM surface morphology after the ∅4 mm projectile impact test. (a), (b), (c)** It shows the graphitic layers from the impact surface of the shattered disk (the side on which the projectile impacted). **(d), (e), (f)** It shows the graphitic layer from the fractured lateral edge of the disk. Images from both sides undoubtably confirm the formation of graphitic layers throughout the bulk of the composite, after the ∅4 mm projectile impact (as similarly seen through XRM imaging).

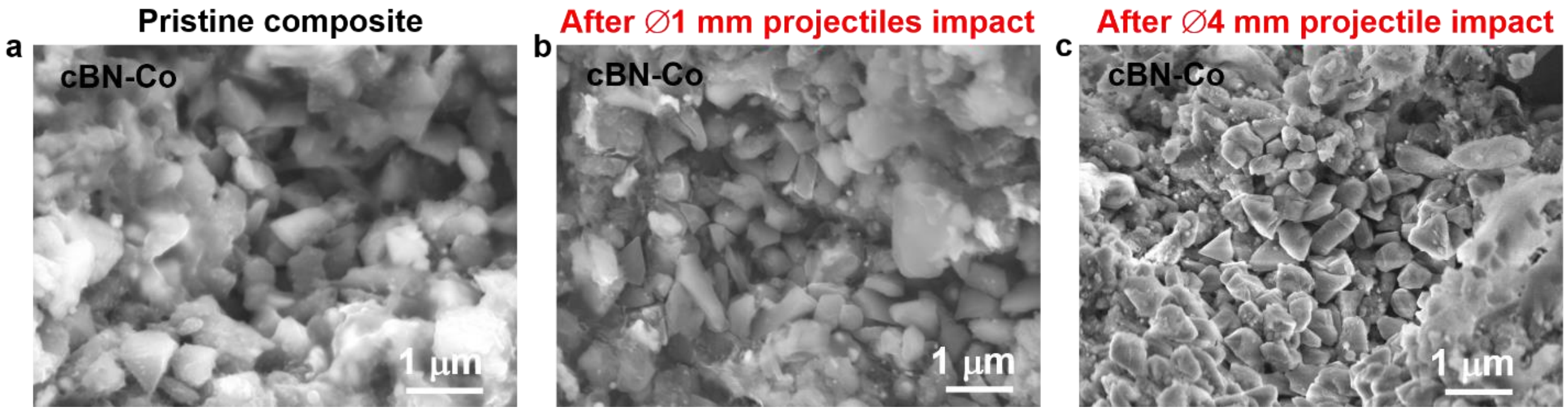

**Fig. S20. Presence of cBN in diamond-cBN-Co composites.** FESEM morphology in **(a)** pristine and **(b), (c)** after impact composites showing cBN particles, as similarly for the pristine cBN powder (as shown in **Fig. S1b**).

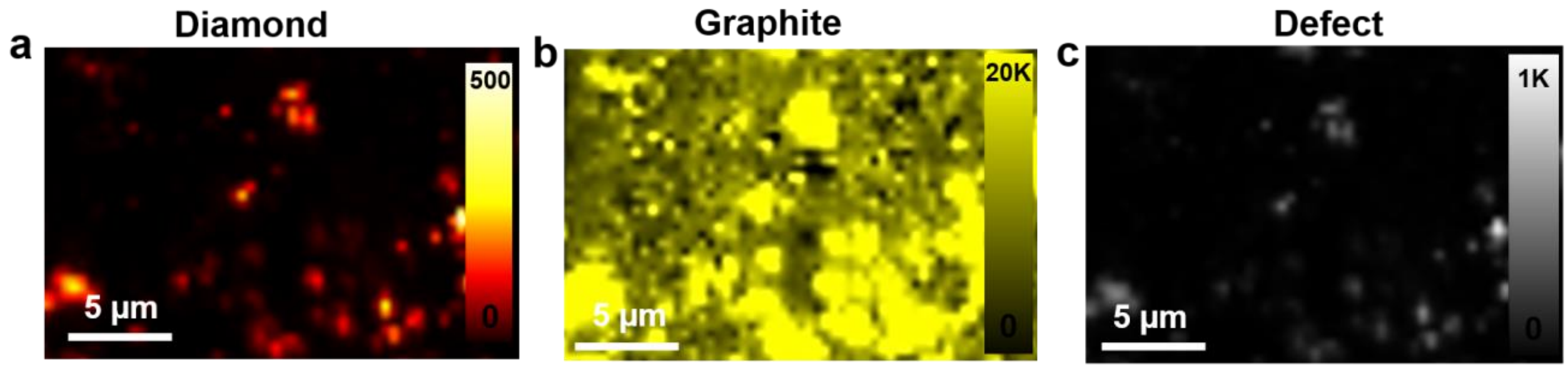

**Fig. S21. Raman intensity map after the ∅4 mm projectile impact test. (a), (b), (c)** Raman intensity map from a graphitic region shows almost no signals for diamond (mostly background), strong graphite peak intensity, and some scattered spots related to defects peak after the impact.

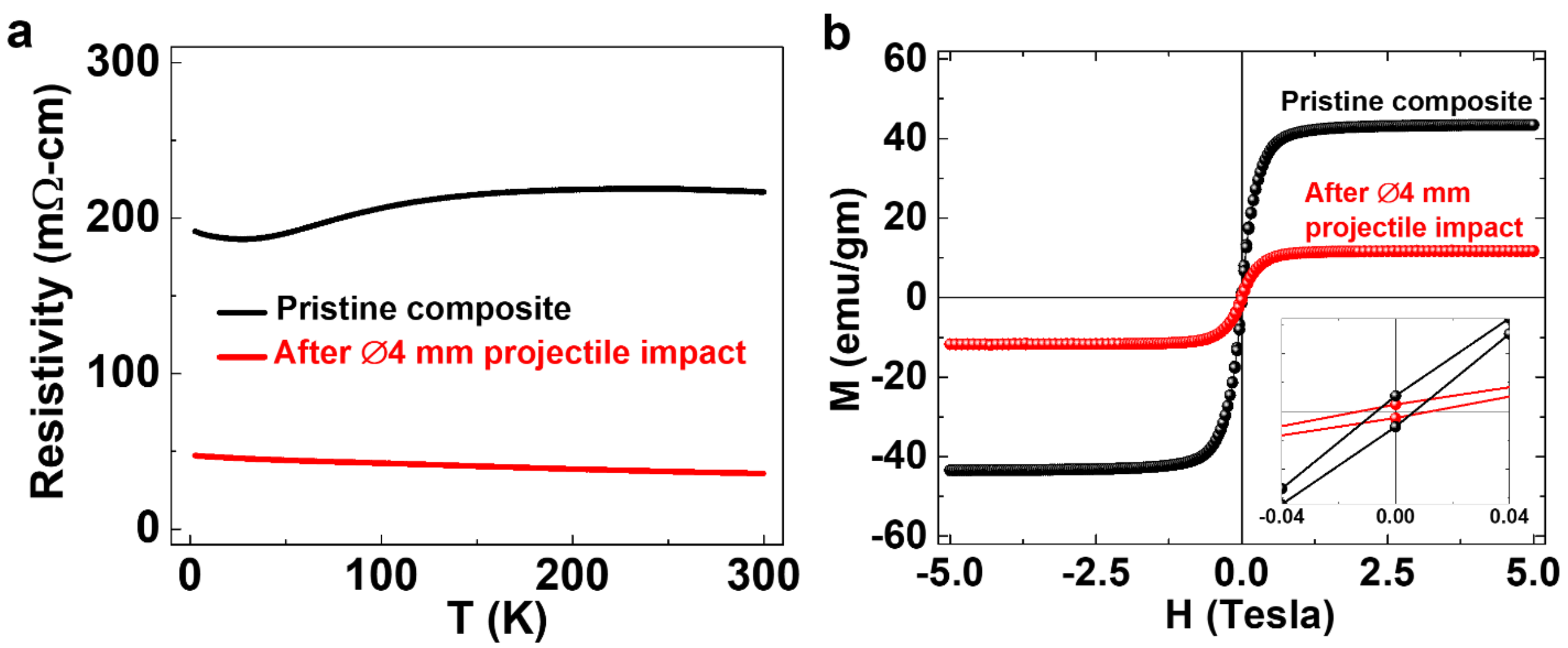

**Fig. S22. Electrical transport and magnetic properties of the composite after the ∅4 mm projectile impact test. (a)** Considering the more graphitic nature of the composite, electrical resistivity decreased with a value of ~36 mΩ-cm at room temperature. **(b)** It still retains a ferromagnetic behavior at room temperature, however with reduced magnetic moments.

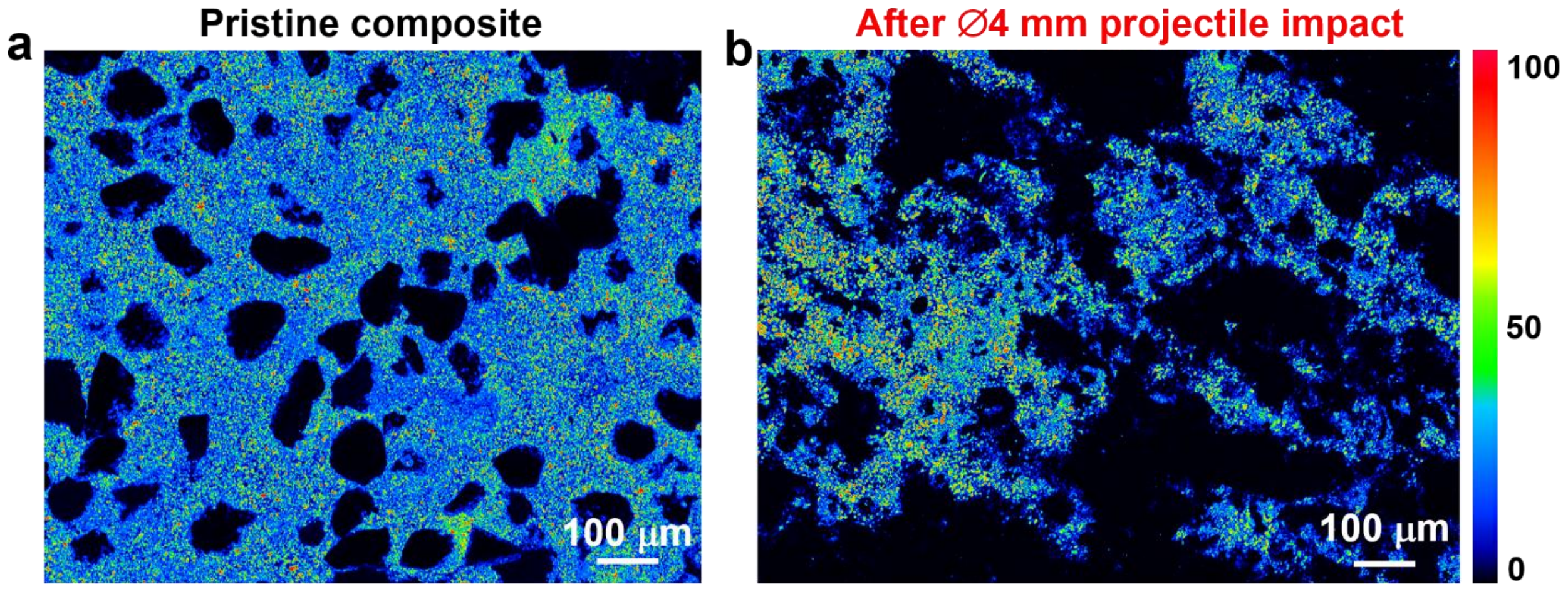

**Fig. S23. WDS elemental map of Co in the ∅4 mm projectile-impacted composite. (a), (b)** Within the random region of analysis, we found Co ~30.88 wt% in the pristine composite, whereas Co ~13.62 wt% in the after ∅4 mm impact sample.

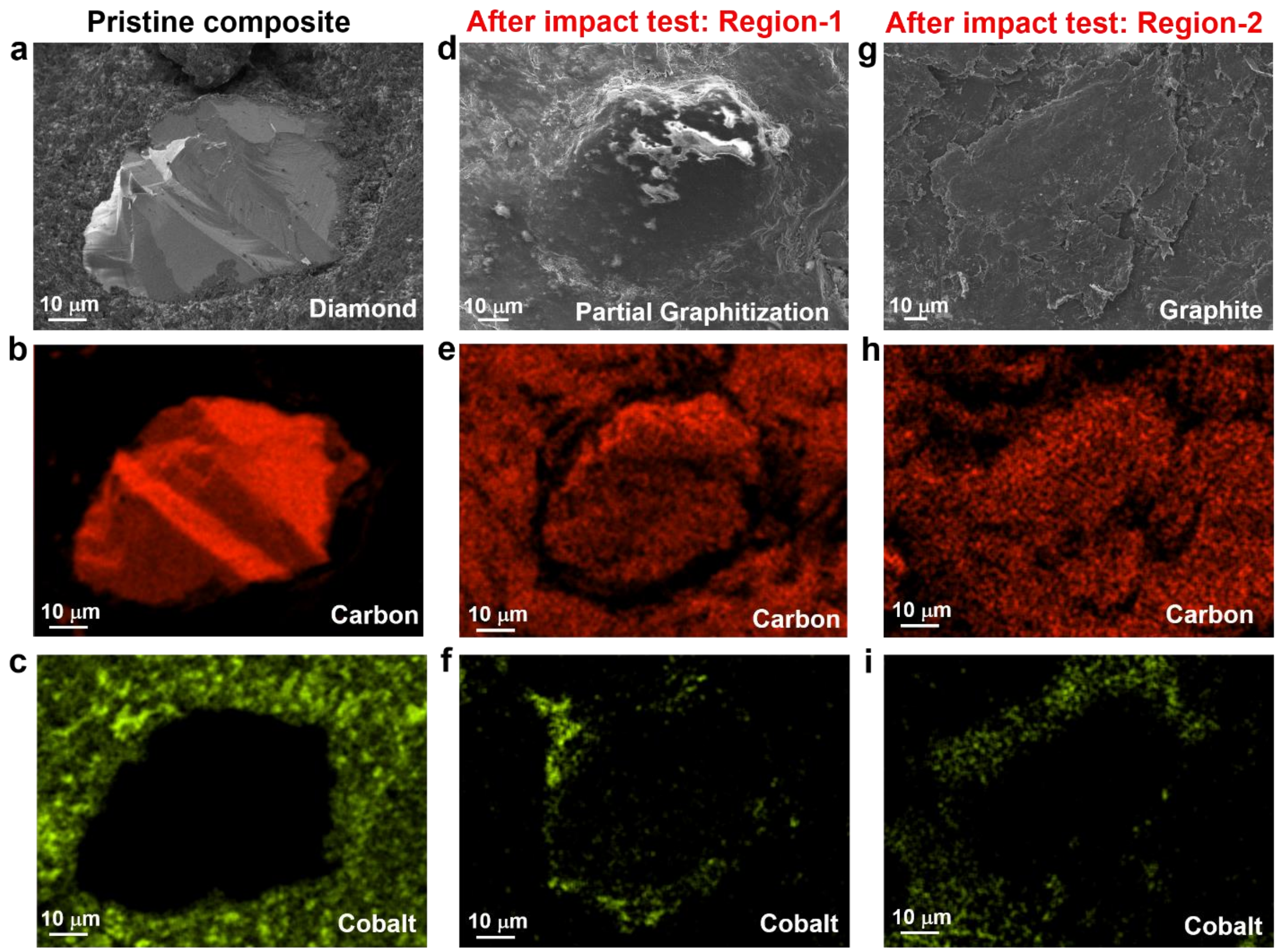

**Fig. S24. Morphological comparison of the diamond particle in the pristine composite and after the ∅4 mm projectile impact. (a), (d), (g)** FESEM surface morphology of a typical diamond particle, which shows that in some cases it's partially graphitized (forming a diamond-graphite region) whereas in some cases it forms fully graphitic layers. (**b**)**, (e), (h)** Corresponding EDS maps showing carbon. **(c), (f), (i**) Co particles, surrounded by the diamond/graphitic regions.

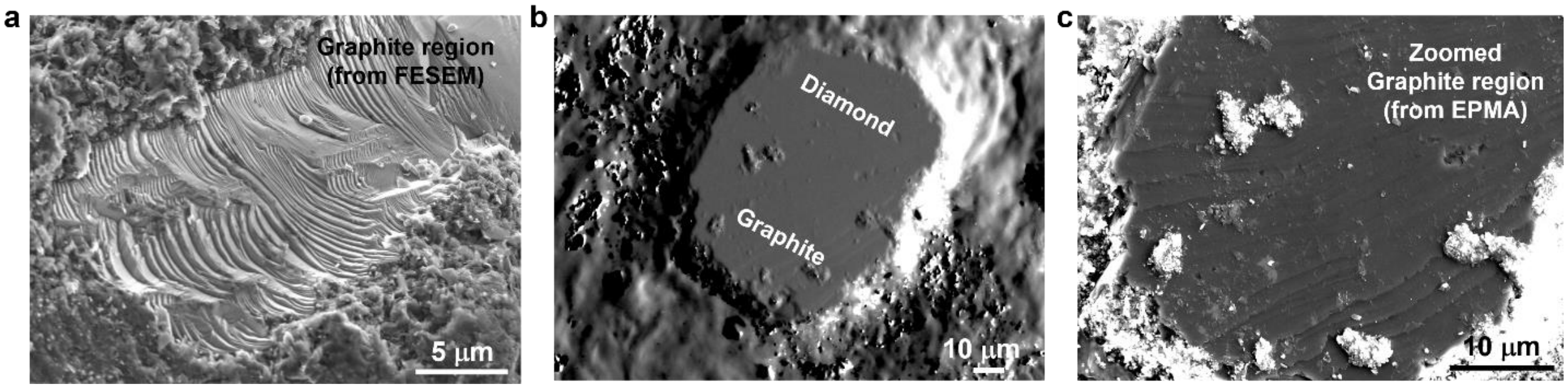

**Fig. S25. Zoomed-in morphology of random diamond grains in the composite after the ∅4 mm projectile impact test.** Zoomed-in **(a)** FESEM morphology, and **(b), (c)** EMPA topography of the diamond grain, showing the formation of layered sheet-like texture at lower regions (zoomed in lower graphitic region in **b**).

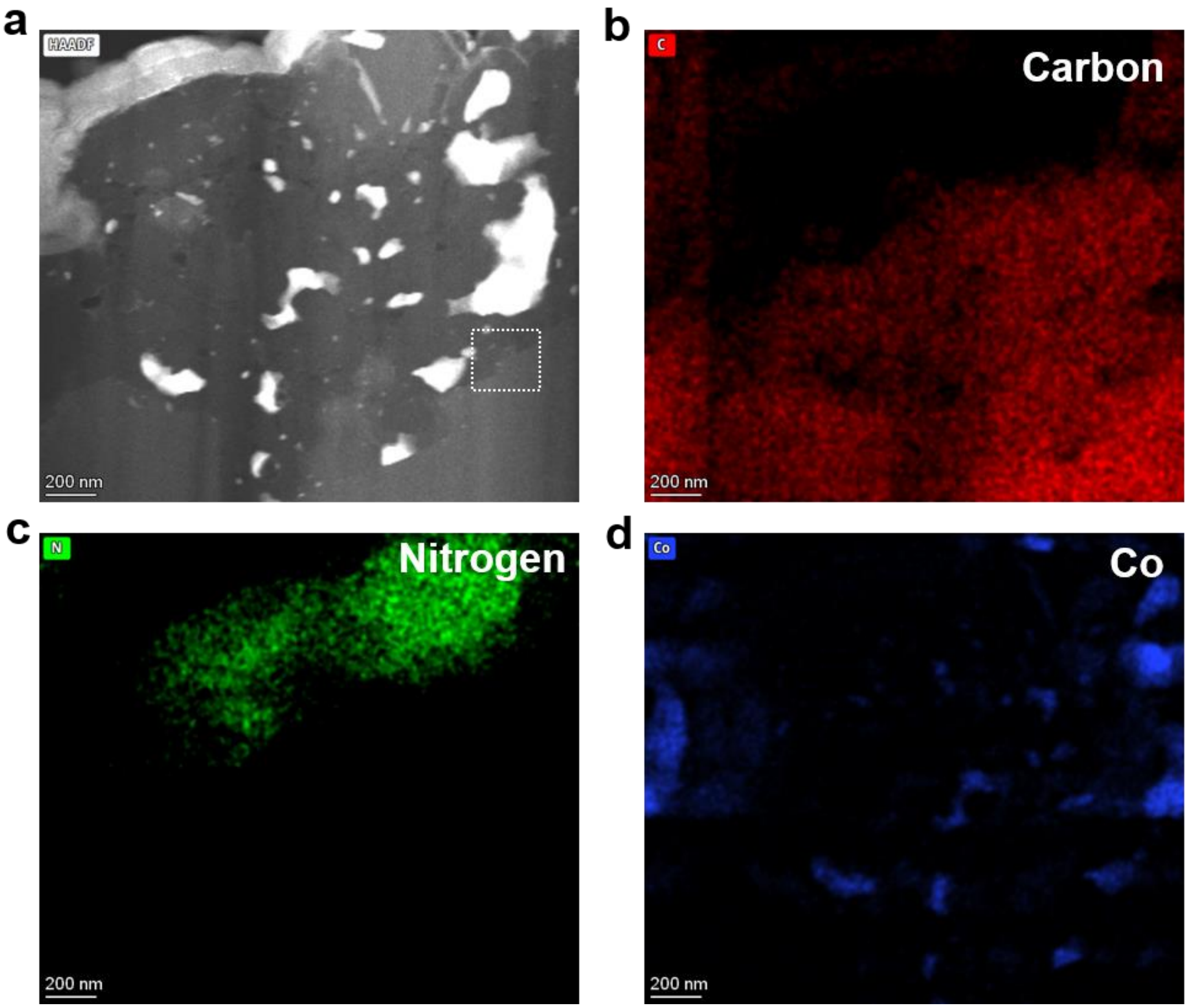

**Fig. S26. Chemical composition mapping by STEM EDX. (a)** Cross-sectional HAADF-STEM image of the after-impact sample. White dashed box is the region where the atomic-scale interface image is shown in the main. **(b), (c), (d)** EDX mapping showing the presence of C, N, and Co.

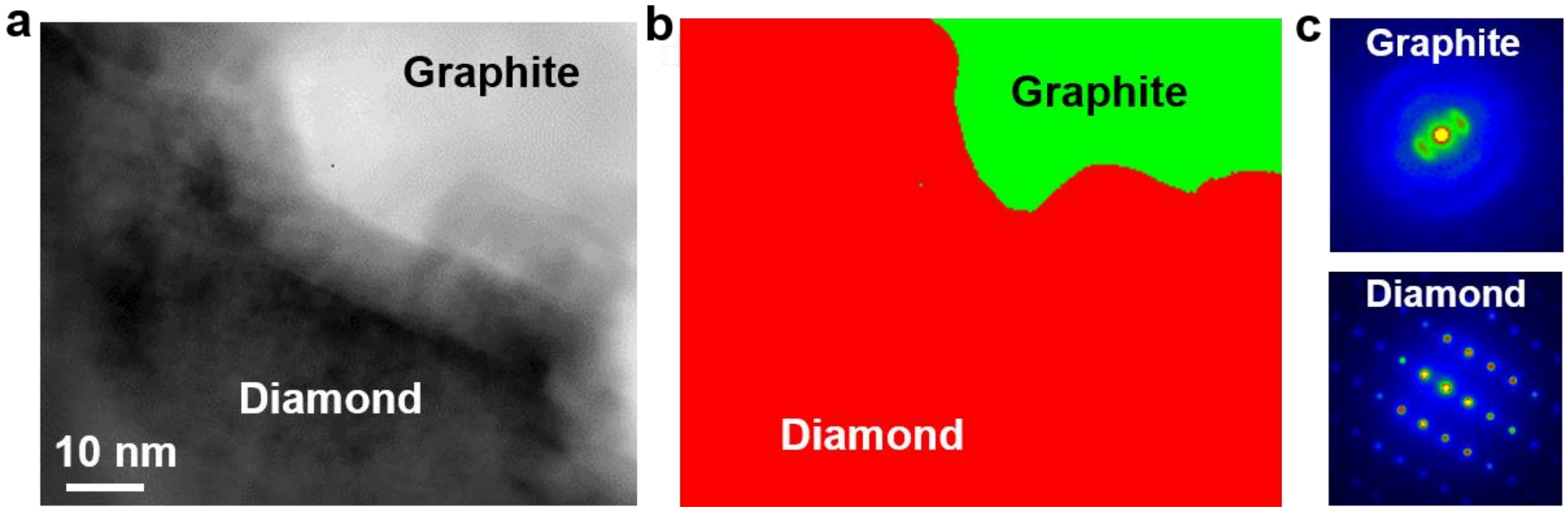

**Fig. S27. Interface between diamond and graphite. (a)** Virtual Bright-field STEM image of diamond-graphite interface. **(b), (c)** Precession 4D-STEM phase mapping and corresponding diffraction patterns from the respective region showing [211] and [110] orientations of diamond and graphite.

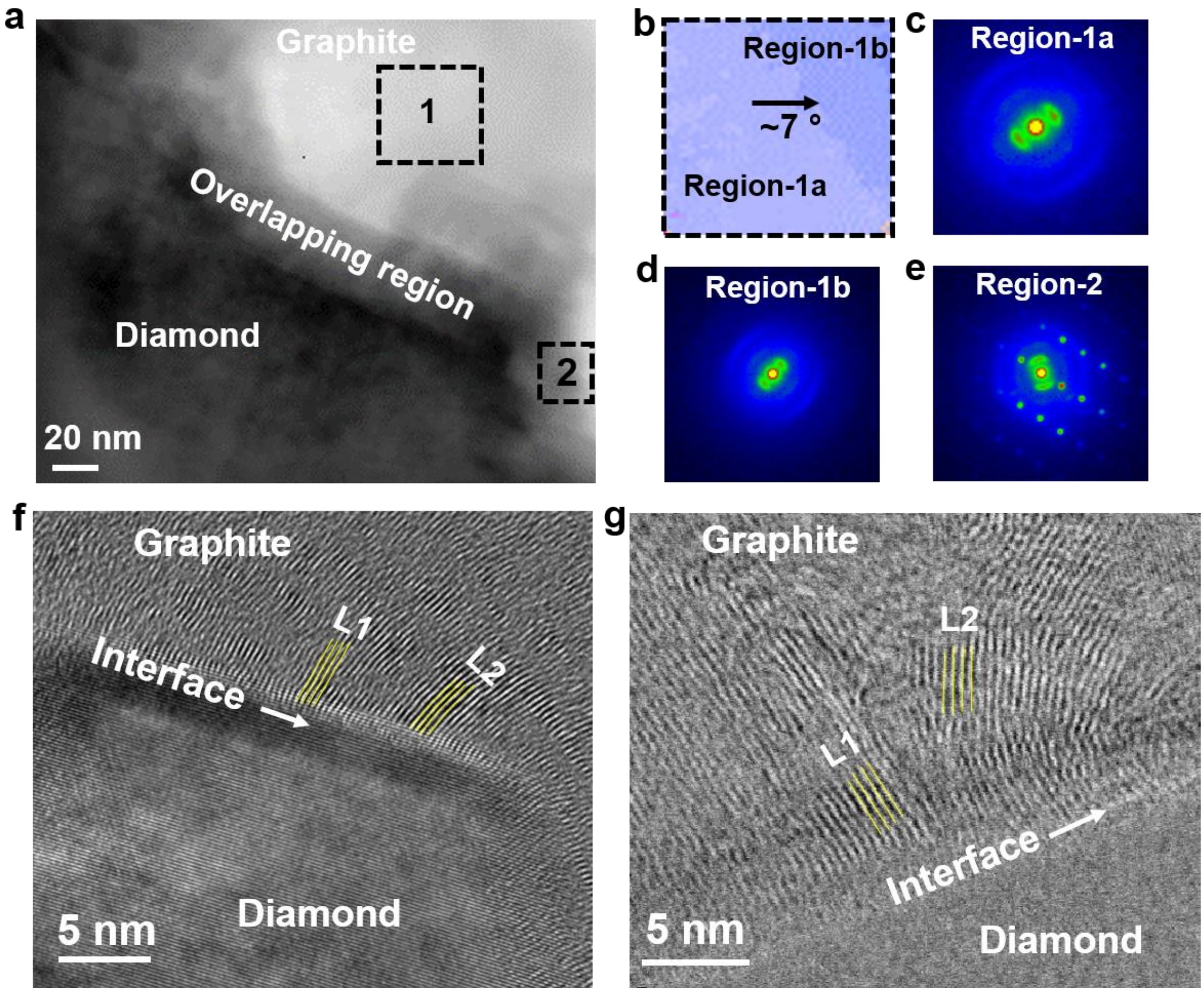

**Fig. S28. 4D-STEM of the diamond-graphite interface. (a)** Virtual bright-field STEM image of the diamond-graphite interface and the overlapped regions. (**b**) Within the black dashed region, it reveals misorientation mapping of graphite layers (~7°). Corresponding 4D STEM diffraction patterns are shown **(c)** from Region-1a, **(d)** from Region-1b, and **(e)** from Region-2. Region-2 shows the presence of both diamond and graphite which is the overlapped region. **(f)** HRTEM shows ~7° misoriented graphitic layers starting from layer L1 to layer L2 (yellow lines), with some defects in the form of dislocations and stacking faults (right bottom). **(g)** ABF STEM shows other misoriented graphitic layers with an angle between the layer L1 and L2 of 32°, from a random interface region.

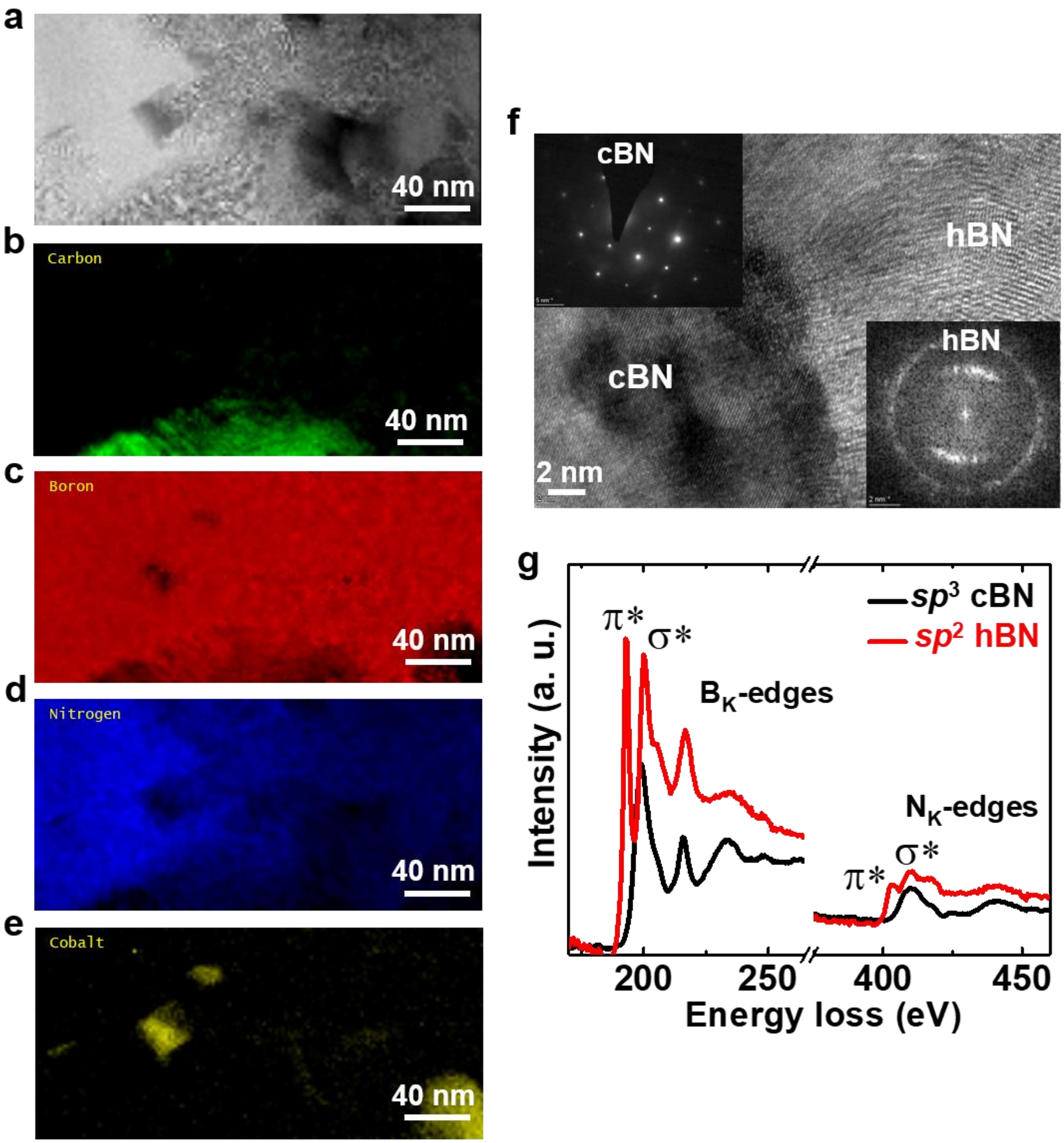

**Fig. S29. HRTEM imaging and STEM EELS of the BN region in the composite after the ∅4 mm projectile impact.** **(a)** ADF STEM image, and **(b), (c), (d), (e)** corresponding EELS map showing the presence of C, B, N, and Co. **(f)** HRTEM along with the FFTs showing the presence of cBN and hBN. **(g)** STEM-EELS spectra from the $sp^3$ cBN and $sp^2$ hBN regions.

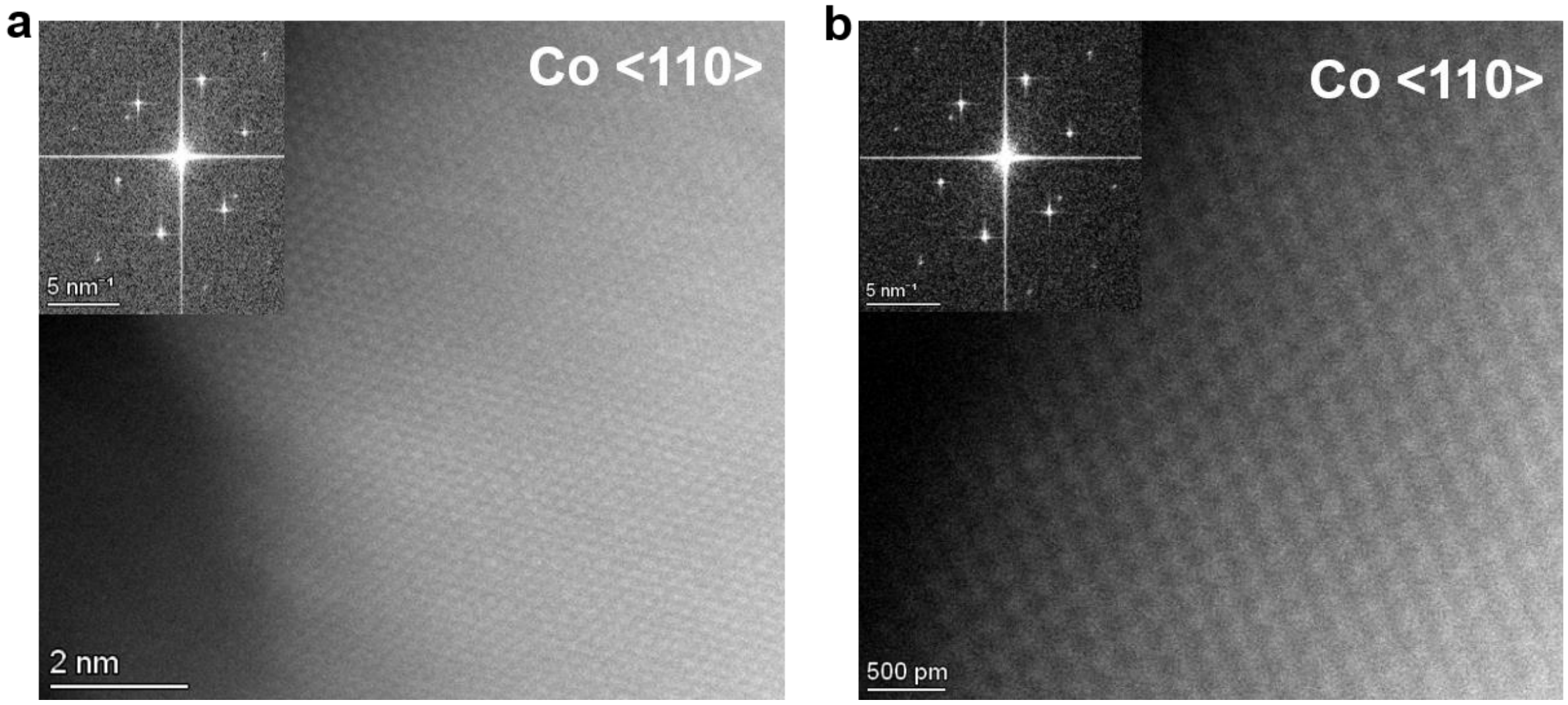

**Fig. S30. TEM images of Co in the composite after the ∅4 mm projectile impact test. (a), (b)** HR-STEM and the FFTs pattern showing <110> orientation of cubic Co lattice from different region of the sample.

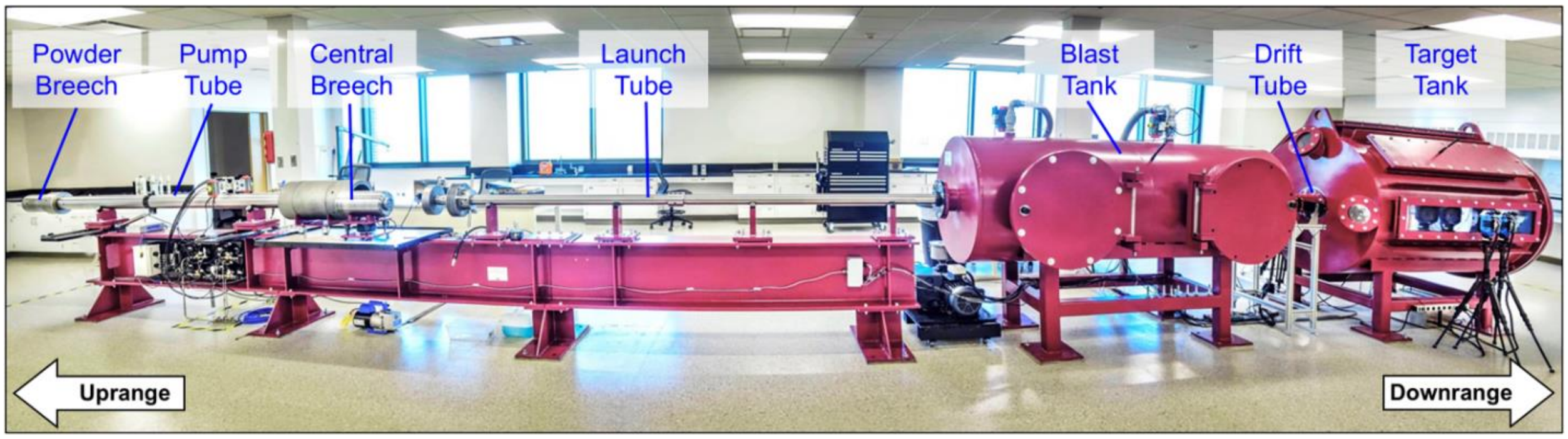

**Fig. S31. Hypervelocity impact test facility**. An overview of the two-stage light-gas gun (2SLGG) at the Texas A&M University (TAMU, USA) Hypervelocity Impact Laboratory (HVIL). The key components include: gun powder breech, pump tube, central breech, nominally 12.7 mm diameter smooth-bore launch tube, blast tank, drift tube with laser velocimetry system, and target tank. The gunpowder loading occurs at the uprange end, while the projectile impacts the composite sample in the target tank located downrange.

## Supplementary Tables

**Table S1**. **Elements analyzed by EPMA**: Spectrometer conditions: X-ray lines, standards used, peak position, and background offsets

| **Element** | **X-ray** | **Analyzing Crystal** | **Standard** | **Peak Pos. (L-value) (mm)** | **Lower Backgr. offset (mm)** | **Upper Backgr. offset (mm)** |
|---|---|---|---|---|---|---|
| C | $K_\alpha$ | LDE2 | C_graphite_SPI | 127.464 | 6 | 5 |
| B | $K_\alpha$ | LDE2 | BN_SPI | 194.516 | 6 | 5 |
| N | $K_\alpha$ | LDE1 | BN_SPI | 147.201 | 5 | 5 |
| Co | $K_\alpha$ | LIFL | Co_SPI | 124.515 | 5 | 5 |

**Table S2. Detector conditions for EPMA analysis**: counting time (Peak=$P_k$ and background = $B_k$) and detector conditions.

| **Element** | **$P_k$ counting (sec)** | **$B_k$ counting (sec)** | **Peak search** | **Mass (%)** | **Detector conditions** | | | | |
|---|---|---|---|---|---|---|---|---|---|
| | | | | | **Gain** | **High.V (V)** | **Base Line (V)** | **Window (V)** | **Mode** |
| C | 10 | 5 | Yes | 99.99 | 64 | 1700 | 0.7 | 9.3 (V) | Dif |
| B | 10 | 5 | Yes | 43.6 | 64 | 1700 | 0.7 | 9.3 (V) | Dif |
| N | 10 | 5 | No | 56.44 | 64 | 1645 | 1.4 | 8.7 (V) | Dif |
| Co | 10 | 5 | Yes | 99.995 | 16 | 1612 | 0.7 | 0 | Int |

**Supplementary Text**

To identify the hexagonal regions (**Fig. 2k**) formed during graphitization induced by projectile impact, a geometric-topological procedure was applied to the configurations of the MD simulation. Atomic bonds were defined exclusively using a geometric criterion, in which two atoms were considered bonded if their interatomic distance was smaller than a cutoff value fixed at 1.7 Å. Based on this criterion, an undirected atomic connectivity graph was constructed, with vertices representing atoms and edges corresponding to local bonds. To distinguish graphitic domains from remnant diamond structures, a topological filter based on atomic coordination was applied, excluding from the ring analysis all atoms with coordination numbers greater than three, which are characteristic of the tetrahedral $sp^3$ bonding network of diamond.

Ring detection was then performed on the resulting subgraph containing only atoms with coordination numbers less than or equal to three. Simple cycles of length six were explicitly identified using a depth-first search (DFS) algorithm constrained to closed paths composed of six distinct vertices, ensuring the detection of genuine hexagonal rings typical of graphitic structures. Additional pruning strategies were employed to prevent redundant detections by enforcing an ordering constraint on the initial vertices of the cycles. All atoms participating in at least one identified six-membered ring were classified as belonging to hexagonal regions, whose spatial distribution is highlighted (**Fig. 2k**). The same procedure was applied to every frame of the trajectory and extended to closed cycles of five and seven vertices, giving the ring populations reported (**Fig. S15**). The bond-angle distribution and the radial distribution function were computed directly from the atomic positions. All populations are referred to the instant of contact, which is taken as the zero of the time axes and of the ring counts. Ring statistics are a standard descriptor of medium-range order in network solids [**S1**]. We note that the present analysis enumerates closed simple cycles of a given length, rather than primitive rings defined by a shortest-path criterion, and the reported counts should be interpreted accordingly.

[**S1**] D. S. Franzblau, Computation of ring statistics for network models of solids, Phys. Rev. B 44, (1991) 4925.

**Supplementary Movies**

(We provide separate mp4 video files)

**Movie S1. X-ray microcopy of pristine diamond-cBN-Co composite,** Bulk x-ray microscopy (XRM) image of the diamond (red), cBN (white) and cobalt (green) in the diamond-cBN-Co three-phase composite.

**Movie S2. X-ray microcopy of pristine diamond-Co composite,** Bulk XRM image of diamond particles (red) in the diamond-Co composite, without using cBN.

**Movie S3. Impact test with ∅1 mm metal projectile,** Impact test conducted on the composite with several ∅1 mm metal projectile. The time scale of the video is a few microseconds.

**Movie S4. Impact test with a ∅4 mm metal projectile,** Impact test conducted on the composite with a ∅4 mm metal projectile. The time scale of the video is a few microseconds.

**Movie S5. Impact induced $sp^2$ bond formation from simulations,** It shows the formation of a hexagonal graphitic $sp^2$ carbon structure due to impact.

**Movie S6. X-ray microcopy of the composite after the ∅4 mm metal projectile impact test,** XRM image of graphite (black) and diamond grains (red) in the composite after the ∅4 mm projectile impact test.

**Movie S7. Early-stage graphitization by using MD simulations with a reactive force-field (ReaxFF),** It shows the atomic-scale phase transformation of diamond-to-graphite due to impact.